\documentclass[nohyper,11pt,letterpaper]{JHEP3}
\usepackage[dvips]{epsfig}

\newcommand{\rp}{r_p}
\newcommand{\vr}{\varrho}
\newcommand{\bm}{\bar{M}}
\newcommand{\gym}{g_{YM}}
\newcommand{\Rii}{R_{11}}
\newcommand{\xii}{x_{11}}

\newcommand{\beq}{\begin{equation}}
\newcommand{\eeq}{\end{equation}}
\newcommand{\beqa}{\begin{eqnarray}}
\newcommand{\eeqa}{\end{eqnarray}}
\newcommand{\beqar}{\begin{eqnarray*}}
\newcommand{\eeqar}{\end{eqnarray*}}

\newcommand{\labell}[1]{\label{#1}} 
\newcommand{\reef}[1]{(\ref{#1})}
\newcommand{\ssc}{\scriptscriptstyle}
\newcommand{\eg}{{\it e.g.,}\ }
\newcommand{\ie}{{\it i.e.,}\ }

\newcommand{\norm}[1]{\raise.3ex\hbox{:}#1\raise.3ex\hbox{:}}

\newcommand{\al}{\alpha}

\newcommand\E{{\cal E}}

\newcommand\lp{{\ell_p}}  

\newcommand\gs{g_s} 
\newcommand\gef{g_{eff}} 

\newcommand\x{\times}
\newcommand\lap[1]{\nabla^2_{\!\!\ssc [#1]}\,}

\title{Holographic mesons in various dimensions}

\author{Robert C. Myers$^{a,b}$ and Rowan M. Thomson$^{a,b}$\\
$^a$ Perimeter Institute for Theoretical Physics \\
Waterloo, Ontario N2L 2Y5, Canada \\
$^b$ Department of Physics, University of Waterloo  \\
Waterloo, Ontario N2L 3G1, Canada \\E-mail:
\email{rmyers@perimeterinstitute.ca, rthomson@perimeterinstitute.ca
}}

\abstract{Following the analysis of \cite{spectra}, we calculate the
spectrum of fluctuations of a probe D$k$-brane in the background of
$N$ D$p$-branes,  for $k=p,p+2,p+4$ and $p< 5$. The result
corresponds to the mesonic spectrum of a $(p+1)$-dimensional
super-Yang-Mills (SYM) theory coupled to `dynamical quarks', \ie
fields in the fundamental representation -- the latter are confined
to a defect for $k=p$ and $p+2$. We find a universal behaviour where
the spectrum is discrete and the mesons are deeply bound. The mass
gap and spectrum are set by the scale $M\sim m_q/\gef(m_q)$, where
$m_q$ is the mass of the fundamental fields and $\gef(m_q)$ is the
effective coupling evaluated at the quark mass, \ie
$\gef^2(m_q)=\gym^2N\,m_q^{p-3}$. We consider the evolution of the
meson spectra into the far infrared of three-dimensional SYM, where
the gravity dual lifts to M-theory. We also argue that the mass
scale appearing in the meson spectra is dictated by holography.}


\preprint{}

\begin{document}{\vskip 1cm}


\section{Introduction}

The AdS/CFT correspondence and its extensions have provided a
powerful framework for the study of strongly coupled gauge theories
in various dimensions. The original correspondence
\cite{malda, adscft}, that certain conformal field theories are
equivalent to string theory on AdS backgrounds, was extended in
\cite{itzhaki} to a more general gauge/gravity duality in various
dimensions. Itzhaki {\it et al} \cite{itzhaki} studied the general
case of a stack of $N$ coincident D$p$-branes in the limit in which
brane modes decouple from the bulk.  They argued that the
super-Yang-Mills $U(N)$ gauge theory on the ($p+1$)-dimensional
worldvolume of the D$p$-branes is dual to the closed string theory
on the `near-horizon' background induced by the branes.

These dualities follow from a straightforward extension of the
usual decoupling limit \cite{bigRev} now applied to D$p$-branes
\cite{itzhaki}. For general $p$ ($\ne3$), the gauge theory is
distinguished from the conformal case by the fact that the
Yang-Mills coupling $\gym$ is dimensionful. Hence there is a
power-law running of the effective coupling with the energy scale
$U$:
\beq g_{eff}^2 = {g_{YM}^2N}\,U^{p-3}\ .\labell{couple}
\eeq
The duality relates the energy scale and the radial coordinate $r$
transverse to the D$p$-brane worldvolume in the usual way,
$U=r/\al'$. In the dual background, the absence of conformal
invariance in the general case is manifest in the radial variation
of both the string coupling (or dilaton) and the spacetime curvature
--- see below for details. As the supergravity background is only
trustworthy for weak string coupling and small curvatures, it
provides a dual description of the theory which is reliable for an
intermediate regime of energies. In this regime, the dual gauge
theory is always strongly coupled.

In a complementary direction, Karch and Katz \cite{karch}
demonstrated that probe D7-branes can be used to introduce
fundamental matter fields into the standard AdS/CFT correspondence
from D3-branes. Inserting $n$ D7-branes into the AdS$_5\times S^5$
background corresponds to coupling $n$ flavours of `dynamical
quarks' (\ie $n$ hypermultiplets in the fundamental representation)
to the original four-dimensional SYM theory. Adding these extra
branes/fields also reduces the number of conserved supercharges from
sixteen to eight. The hypermultiplets arise from the lightest modes
of strings stretching between the D$7$- and D$3$-branes and their
mass is $m_q = L/2\pi \al '$ where $L$ is the coordinate distance
between the two sets of branes (and as usual, $1/2\pi \al '$ is the
string tension). The resulting gauge theory containing quarks has a
rich spectrum of quark-antiquark bound states, which henceforth we
refer to as `mesons' \cite{spectra}. In the decoupling limit, the
duals are open strings attached to the D$7$-branes and the
calculation of the meson spectrum in the field theory becomes an
exercise in studying the fluctuation of probe branes. These ideas
have been further developed in a number of directions towards the
goal of constructing gauge/gravity duals for a QCD-like theories
\cite{holoQCD,quarkrefs,sakai} and in particular, the meson spectrum
has been studied in a number of different contexts
\cite{spectra,mesonrefs}.

In this paper, we use the gauge/gravity duality to explore the meson
spectra of such SYM theories containing fundamental fields in
different numbers of spacetime dimensions. In particular, following
\cite{spectra}, we calculate the spectrum of fluctuations of a probe
D$k$-brane supersymmetrically embedded in the background of $N$
D$p$-branes, for $k=p,p+2,p+4$ and $p<5$. This corresponds to the
mesonic spectrum of a $U(N)$ super-Yang-Mills theory in $p+1$
dimensions coupled to a hypermultiplet in the fundamental
representation. Again, these dynamical quarks arise as the lightest
modes of the $(k,p)$ and $(p,k)$ with their mass given by $m_q =
L/2\pi \al '$. Also as before, the number of conserved supercharges
is reduced to eight by the addition of the fundamental
hypermultiplet. A consequence of the supersymmetric embedding of the
probe brane is that the quarks are confined to a defect of
codimension two and one for $k=p$ and $p+2$, respectively. So it is
only for $k=p+4$ that the D-brane configuration yields fundamental
fields propagating in the full $p+1$ dimensions of the gauge
theory.\footnote{However, for the $k=p$ and $p+2$ cases, one might
consider compactifying the D$p$-brane worldvolume directions
transverse to the D$k$-brane, as in \cite{holoQCD}.}

The resulting meson spectra for all of these different
configurations display certain universal characteristics, which
are common to the original D3-D7 results \cite{spectra}. In
general, the spectra are discrete and the mesons are deeply bound.
Up to numerical coefficients, the meson masses can all be written
in terms of a single scale:
\beq M\sim m_q/\gef(m_q)\ , \labell{massgap} \eeq
where $m_q$ is the quark mass and $\gef(m_q)$ is the effective
coupling \reef{couple} evaluated at this mass scale,
$\gef^2(m_q)=\gym^2N\,m_q^{p-3}$. In particular then,
eq.~\reef{massgap} gives the mass gap of the spectrum. The detailed
calculation of these results is presented in section \ref{sugraM}.

As an example, for $N$ background D2-branes and a D6-brane probe, we
find that the meson masses scale as $M \propto m_q^{3/2}/(\gym^2
N)^{1/2}$.  It is interesting to note then that in \cite{ek}, a
similar study found $M \propto m_q$ for the same configuration of
branes. The resolution of this apparent discrepancy is that the
latter analysis considers very low quark masses in the far infrared,
\ie $m_q\ll\gym$, where the dual gravity configuration actually
lifts to M-theory. To better understand this behaviour in different
regimes of the field theory, in section \ref{beyond}, we study probe
D2- and D4-branes in the D2-brane background, for which computations
can easily be extrapolated between the type IIA and M-theory
regimes. Hence, in this section, we calculate the meson spectra in
$(2+1)$-dimensional SYM coupled to a fundamental hypermultiplet
confined to codimension-two and -one defects, in the far infrared
regime of the theory.

Finally, section \ref{discus} presents a discussion of our results
and future directions. In particular, we present a simple argument
that the mass scale of the meson spectra is dictated by the
consistency of holography and we briefly describe several tests of
the latter idea. A number of technical results are provided in
subsequent appendices. While this paper was in preparation, we
became aware of \cite{spain} which addresses the same basic problem
as considered here, however, we provide some complementary analyses
and a different interpretation of the results.


\section{Supergravity meson spectra in various dimensions}\label{sugraM}

In this section, we study the excitations of a D$k$-brane probe
(with $k=p+4,p+2,p$) supersymmetrically embedded in the `near
horizon' geometry induced by $N$ coincident D$p$-branes (with
$p<5$).\footnote{There is no dual field theory for $p\ge6$ as no
decoupling limit is possible \cite{itzhaki,seiberg}. We comment on
the interesting case of $p=5$ in section \ref{discus}.} These
calculations yield the spectrum of mesonic states in the dual
$(p+1)$-dimensional field theory with eight supercharges. We first
review the background geometry induced by the D$p$-branes and then
proceed to compute the spectra of fluctuations of the probe
D$k$-branes.


\subsection{Background geometry}

The supergravity solution corresponding to $N$ coincident
D$p$-branes is, in the string frame (see, \eg \cite{johnson} and
references therein)
\beqa
ds^2 &=& f_p^{-1/2} ds^2(\mathbb{E}^{(1,p)}) + f_p^{1/2}
d\vec{Y} \cdot d\vec{Y} \nonumber \\
e^{2\phi} &=&  f_p^{\,(3-p)/2}\qquad\qquad C_{0...p} =
-(f_p^{-1}-1)\labell{fullGeom} \eeqa
where $ds^2(\mathbb{E}^{(1,p)})$ denotes a flat spacetime metric
with one time and $p$ spatial directions.   The $Y^A$, with
$A=1,...,9-p$, parametrize the $(9-p)$-dimensional space
transverse to the D$p$-branes.\footnote{These directions
correspond to $x^A$ with $A=p+1,\,p+2,\,\cdots,\,9$ in the arrays
presented below --- see, \eg eq.~\reef{Dp-D(p+4)}.} The harmonic
function $f_p$ depends on the transverse radial coordinate
$r=|\vec{Y}|$
\beq f_p = 1+ \left(\frac{r_p}{r}\right)^{7-p} \labell{harm0}\eeq
where the constant $\rp$ is defined in terms of the number $N$ of
D$p$-branes, the string coupling constant $\gs$, and the inverse
string tension $\al'$:
\beq  r_p^{7-p}=  \gs N\,
(4\pi\al')^{(7-p)/2}\,\Gamma\!\left({\scriptstyle\frac{7-p}{2}}\right)
/4\pi\ . \labell{rp}\eeq
Following \cite{itzhaki}, we take the decoupling limit (in which
the open string modes propagating on the D$p$-branes decouple from
the bulk closed string modes) with
\beq  \gym^2 = (2\pi)^{p-2}\gs\, \al'^{(p-3)/2} = fixed, \qquad \al'
\to 0. \labell{gym}\eeq
This limit also holds $r/\al' $ constant \cite{malda} so that
eq.~\reef{fullGeom} reduces to the `near horizon' solution:
\beqa  ds^2 &=& \left( \frac{r}{r_p}\right)^{(7-p)/2}
ds^2(\mathbb{E}^{(1,p)}) + \left( \frac{r_p}{r}\right)^{(7-p)/2}
d\vec{Y} \cdot d\vec{Y} \nonumber\\
e^{2\phi} &=& \left( \frac{\rp}{r}\right)^{(7-p)(3-p)/2}\qquad
C_{0...p} = -\left(\frac{r}{r_p}\right)^{7-p}\ .\labell{nearGeom}
\eeqa
This supergravity background then provides a dual description of
$U(N)$ super-Yang-Mills theories in $p+1$ dimensions, the
worldvolume field theory on the D$p$-branes. In accord with the
duality, both the background and the field theory have sixteen
supersymmetries.  The isometry group for the background geometry
\reef{nearGeom} induced by the D$p$-branes is $SO(1,p) \times
SO(9-p)$ \cite{itzhaki}.  From the perspective of the dual gauge
theory, $SO(1,p)$ corresponds to the spacetime Lorentz symmetry
while $SO(9-p)$ is the R-symmetry group.

As mentioned in the introduction, the supergravity solution
\reef{nearGeom} is a trustworthy background provided that both the
curvatures and string coupling are small. This limits the
supergravity description to an intermediate regime of energies in
the field theory or of radial distances in the background. In terms
of the effective coupling \reef{couple}, this restriction is
succinctly expressed as \cite{itzhaki}
\beq 1\ll \gef\ll N^{\frac{4}{7-p}}\ .\labell{range}\eeq
Hence the field theory is strongly coupled where the dual
supergravity description is valid. For $p<3$, the spacetime
curvature is large for very large values of $r$, invalidating the
supergravity description.  However, in this UV regime, the
coupling runs to $\gef\ll 1$ and so perturbative field theory is
applicable. For small values of $r$, both the effective coupling
and the dilaton are large. However, another dual description can
be found to describe this infrared regime with $\gef\gg
N^{\frac{4}{7-p}}$ \cite{itzhaki}. For $p>3$, the coupling runs in
the reverse direction. Hence perturbative SYM applies in the far
infrared where one finds large spacetime curvatures for small $r$.
The effective coupling grows in the UV and a new dual theory must
be found for very large $r$. For $p=3$, the theory is conformal
and the supergravity background \reef{nearGeom} can be used for
all energy regimes \cite{malda}.


\subsection{Meson spectrum in D$p$-D$(p+4)$}\label{mesonsDpDp4}

Consider the following configuration of $N$ coincident D$p$-branes
($0\leq p <5$) and one D$(p+4)$-brane probe, in which the
D$(p+4)$-brane is parallel to the D$p$-branes' worldvolume
directions:
\begin{equation}
\begin{array}{cccccccccccc} & 0 & 1 & \cdots & p & p+1& p+2 &
p+3 & p+4 & p+5 &\cdots & 9\\
Dp & \x & \x & \cdots &  \x &   & & &  &  & & \\
D(p+4) & \x & \x & \cdots & \x & \x  &\x &\x & \x &  &  & \\
\end{array}\labell{Dp-D(p+4)}
\end{equation}
Embedding the D$(p+4)$-brane in the $(p+5)\cdots 9$ plane at
$|\vec{Y}|=L$, the configuration of branes remains supersymmetric
but only half of the original sixteen supercharges are preserved.
Correspondingly, the $SO(9-p)$ symmetry of the background geometry
\reef{nearGeom} is broken to $SO(4)\sim SU(2)_R \times SU(2)_L $
acting on $Y^1Y^2Y^3Y^4$ and for $L\ne 0$ (which is the case of
interest here) $SO(4-p)$ acting in the remaining transverse space
around the D$(p+4)$-brane. The R-symmetry group for fields on the
D$(p+4)$-brane is $SU(2)_R\times SO(4-p)$. The fields can be
classified according to their transformation properties under
$SU(2)_R \times SU(2)_L\times SO(4-p)$.\footnote{We note, however,
that the $SO(4-p)$ properties tend to be less useful. In particular,
this group only really acts for $p\le 2$.}

The induced metric on the D$(p+4)$-brane probe is, from
(\ref{nearGeom}),
\beq \labell{gind} ds^2 = \left( \frac{ \sqrt{\rho ^2
+L^2}}{r_p}\right)^{\frac{7-p}{2}} ds^2(\mathbb{E}^{(1,p)}) + \left(
\frac{r_p}{\sqrt{\rho ^2 +L^2}}\right)^{\frac{7-p}{2}} (d\rho ^2 + \rho
^2 d\Omega_3^2 ) \eeq
where $\rho ^2 = r^2 - L^2 $ and $\Omega _3$ are spherical
coordinates in the $(p+1) \cdots (p+4)$-space.  The dual gauge
theory is $(p+1)$-dimensional $U(N)$ SYM coupled to a hypermultiplet
of matter fields in the fundamental representation.  The quark mass
is the distance $L$ separating the D$(p+4)$-brane from the
D$p$-branes multiplied by the string tension $1/2\pi \alpha'$:
\beq m_q = \frac{L}{2\pi \alpha'}\ .\labell{qmass} \eeq

We wish to calculate the spectrum of mesons corresponding to open
string excitations on the D$(p+4)$-brane  --- the analysis closely
follows that of \cite{spectra}. These mesons are represented by
excitations of the scalar and gauge worldvolume fields.  Their
dynamics is governed by the D$(p+4)$-brane action. The relevant
terms of the latter include the Dirac-Born-Infeld (DBI) action and
only the term with $C_{(p+1)}$ from the Wess-Zumino action (see, \eg
\cite{johnson} and references therein):
\beq S = S_{DBI} - \tau _{p+4} \frac{(2\pi \al')^2}{2} \int
P[C_{(p+1)}]  \wedge F \wedge F \labell{actionDpDp4} \eeq
where
\beq S_{DBI} = -\tau _{p+4} \int d^{p+5}\sigma e^{-\phi}
\sqrt{-\det (P[G]_{ab}+2 \pi \al' F_{ab})}\ . \labell{DBI} \eeq
Here, $\tau _{p+4}= 2\pi/\gs(4\pi^2 \al')^{(p+5)/2}$ is the
D$(p+4)$-brane tension and, as usual, $P$ denotes the pullback of
a bulk field to the probe brane's worldvolume \cite{dielectric}.
The ten-dimensional spacetime metric $G_{ab}$ and the bulk
Ramond-Ramond (RR) form $C_{(p+1)}$ were given in
\reef{nearGeom}.\footnote{The index labelling conventions here and
in the following are:  The indices $a,b,c,\ldots$ denote the
worldvolume directions of the probe brane. Greek indices $\mu,
\nu$ denote probe brane directions parallel to the background
branes. For directions in the probe brane worldvolume orthogonal
to the background branes, we use spherical polar coordinates with
radius $\rho$ and angular coordinate indices denoted by indices
$i,j,k$. Finally, indices $A,B,C,\ldots$ denote background
directions orthogonal to the D$p$-brane worldvolume.
\label{indexConv}}

Note that the action \reef{actionDpDp4} represents the bosonic part
of the full action invariant under eight supercharges and that it
has $SU(2)_R \times SU(2)_L\times SO(4-p)$ symmetry corresponding to
rotations in the transverse space.  The Wess-Zumino term breaks the
symmetry interchanging the $SU(2)_R$ and $SU(2)_L$.  In the dual
gauge theory, this corresponds to the asymmetry of $SU(2)_L$ and
$SU(2)_R$: the former commutes with the supercharges while the
latter, as the R-symmetry group of the theory, does not.

We begin by considering fluctuations in the position of the probe
D$(p+4)$-brane. Working in the static gauge, these are scalar
fluctuations $\chi^A$ about the fiducial embedding:
\[
Y^A = L\,\delta ^A _9 + 2\pi \al' \chi ^A, \quad  A = 5,...,9-p \
.
\]
We only need to retain terms in eq.~\reef{actionDpDp4} to
quadratic order in these fluctuations and so the relevant part of
the Lagrangian density for fluctuations in the scalar fields is
\beq
\mathcal{L} \simeq -\tau_{p+4}\, e^{-\phi} \sqrt{-\det g_{ab}}
\left[1 + \frac{(2\pi\alpha')^2}{2} \left( \frac{r_p}{r}
\right)^{(7-p)/2}g^{cd}\partial_c \chi^A \partial_d \chi^A
\right] \, , \labell{lag}
\eeq
where $g_{ab}$ denotes the induced metric (\ref{gind}) on the
D$(p+4)$-brane.  Summation over the repeated index $A$ is implied.
The factor $e^{-\phi}\sqrt{-\det g_{ab}}$ is independent of the
fluctuations $\chi^A$, which is a reflection of the supersymmetry of
the brane configuration. The latter dictates there be no potential
for the position of the probe brane. Retaining terms only to
quadratic order in the fluctuations, we can drop terms containing
$\chi^A$ from the factor $g^{cd}/r^{(7-p)/2}$ and so we can use
(\ref{gind}) as the induced metric on the probe brane in the final
result.

The equations of motion for each of the $5-p$ fluctuations
resulting from the variation of the Lagrangian density \reef{lag}
are:
\beq
\partial_a \left[ \frac{\rho^3 \sqrt{\det h}}{(\rho^2+L^2)^{(7-p)/4}}
g^{ab}\partial_b\chi\right]=0 \, ,
\eeq
where we have taken $\chi$ to be any one of the $\chi^A$
fluctuations. Also $h_{ij}$ is the metric on the unit three-sphere
which, along with radial coordinate $\rho$, spans the
$(Y^1,...,Y^4)$ directions.  Expanding the equation of motion, we
obtain:
\beq
\frac{r_p^{7-p}}{(\rho^2+L^2)^{(7-p)/2}} \partial_\mu
\partial^\mu \chi + \frac{1}{\rho^3}\partial _\rho (\rho^3
\partial_\rho \chi) + \frac{1}{\rho^2} \lap3 \chi =0
\labell{eom}
\eeq
where $\lap3$ is the Laplacian on the unit three-sphere. Next,
using separation of variables, we write the modes as
\beq \chi = \phi (\rho) e^{ik\cdot x}\, \mathcal{Y}^{\ell_3}(S^3)
\ , \labell{sepVar} \eeq
where $\mathcal{Y}^{\ell_3 }(S^3)$ are spherical harmonics on the
$S^3$ satisfying
\beq \lap3 \mathcal{Y}^{\ell_3} =-{\ell_3} ({\ell_3}+2)\,
\mathcal{Y}^{\ell_3} \, , \labell{S3} \eeq
and transforming in the $(\frac{{\ell_3}}{2},\frac{{\ell_3}}{2})$
representation of $SO(4)=SU(2)_L \times SU(2)_R$.  Then,
substituting (\ref{sepVar}) into (\ref{eom}), and setting
\beq \vr = \frac{\rho}{L} \qquad  \bm^2
=-\frac{k^2\rp^{7-p}}{L^{5-p}}, \labell{mbar}\eeq
we obtain the following equation for $\phi(\rho) = \phi(\vr)$:
\beq
\partial_\vr^2 \phi(\vr) +\frac{3}{\vr}\partial_\vr \phi(\vr) +
\left( \frac{\bm^2}{(1+\vr^2)^{(7-p)/2}}
-\frac{{\ell_3}({\ell_3}+2)}{\vr^2}\right) \phi (\vr)=0\ . \labell{eom2}\eeq
The solution to this differential equation must be real-valued and
regular \cite{spectra}.  The solutions must also be normalizable in
order to be dual to a meson state in the field theory.  Thus, real
solutions were chosen that were regular at the origin.  The
eigenvalues $\bm$ were then determined by requiring that the
solutions were convergent at $\rho \to \infty$. For $p=3$, the
solutions can be determined in terms of hypergeometric functions
\cite{spectra} -- see Appendix \ref{solns}.\footnote{Analytic
solutions can also be found for eq.~\reef{eom2} in the case $p=5$
but one finds there are no normalizable solutions!} However, for
general $p$, there are no known analytic solutions to this equation
and one must resort to numerics.

Solving (\ref{eom2}) to find the eigenfunctions $\phi(\vr)$ and
the dimensionless eigenvalue $\bm$, yields the $(p+1)$-dimensional
mass spectrum of mesons $M^2 = -k^2= \bm^2 L^{5-p}/\rp^{7-p}$. The
mass eigenvalues will depend on the radial quantum number $n$
(which corresponds to the number of nodes in $\phi(\vr)$, the
radial profile) and the angular momentum quantum number $\ell_3$.
Hence we denote the eigenvalues for these scalar fluctuations as
$\bm_s^2(n,\ell_3)$.

Using \reef{gym} and \reef{qmass}, the spectrum of excitations can
be expressed in terms of field theory quantities as follows:
\beq  M^2 = \frac{m_q^{5-p}}{\gym^2 N} \left(
\frac{2^{p-2}\pi^{(p+1)/2}}{\Gamma(\frac{7-p}{2})} \right)\bm^2\ .
\labell{mass}\eeq
Further then from eq.~\reef{couple}, we see that the meson masses
scale a $M \propto m_q / g_{eff}(m_q)$.


Turning to the fluctuations of the gauge fields on the probe
brane, the equations of motion for these fields follow from
\reef{actionDpDp4} as
\beq
\partial_a \left(e^{-\phi} \sqrt{-\det g_{cd}} F^{ab}  \right) -
\frac{7-p}{r_p^{7-p}} \, \rho \, (\rho ^2 +L^2)^{(5-p)/2} \,
\epsilon^{bij} \, \nabla_{[3]i} A_j =0 \, , \labell{gaugeEom}
\eeq
where $\epsilon^{bij}$ is a antisymmetric tensor density on the
three-sphere, taking values $\pm 1$. Also $\nabla_{[3]i} $ is the
covariant derivative on the $S^3$ of unit radius, and coordinates
are indexed as described in footnote \ref{indexConv}.  The first
term comes from the Dirac-Born-Infeld part of the action, while
the second is from the Wess-Zumino term and is only present if $b$
corresponds to an $S^3$ index.

We proceed to solve (\ref{gaugeEom}) by expanding $A_\mu$ and
$A_\rho$ in (scalar) spherical harmonics and $A_i$  in vector
spherical harmonics.  There are three classes of vector spherical
harmonics.  The first is the covariant derivative on the
three-sphere of the usual scalar spherical harmonics, $\nabla
_{[3]i} \mathcal{Y}^{\ell_3} (S^3)$, while the other two, labelled
$\mathcal{Y}^{{\ell_3} ,\pm} _i (S^3)$, have ${\ell_3} \geq 1$,
transform in the $(\frac{{\ell_3} \mp 1}{2}, \frac{{\ell_3} \pm
1}{2})$ of $SO(4)$, and satisfy
\begin{eqnarray}
\nabla^2_{[3]} \mathcal{Y}^{{\ell_3} ,\pm} _j - R^k_j
\mathcal{Y}^{{\ell_3} ,\pm} _k &=& -({\ell_3} +1) ^2
\mathcal{Y}^{{\ell_3} ,\pm} _j, \nonumber \\
\epsilon_{ijk} \nabla_{[3]j} \mathcal{Y}^{{\ell_3} ,\pm}_k
 &=& \pm ({\ell_3} +1) \,\mathcal{Y}^{{\ell_3} ,\pm} _i, \nonumber\\
\nabla^i_{[3]} \mathcal{Y}^{{\ell_3} ,\pm} _i &=& 0 \labell{harm
id}
\end{eqnarray}
where $R^k _j =2\delta ^k _j$ is the Ricci tensor on the
three-sphere of unit radius.

We first consider modes satisfying $\partial_\mu A^\mu =0$.  In
this case, the equation of motion  for $A_\mu$ decouples from the
other equations of motion, so we have the following two types of
modes:
\begin{equation}\label{type 1}
\textrm{Type 1:} \quad  A_\mu =\zeta _\mu \phi_{1}(\rho)
 e^{ik\cdot x}\, \mathcal{Y} ^{\ell_3} (S^3), \quad  k\cdot \zeta = 0,
 \quad A_\rho = 0,  \quad A_i = 0
\end{equation}
and
\begin{equation}\label{type 2}
\textrm{Type 2:} \quad  A_\mu =0,  \quad  A_\rho = \phi_{2}(\rho)
 e^{ik\cdot x}\, \mathcal{Y} ^{\ell_3} (S^3),  \quad A_i = \tilde{\phi}_{2}
  (\rho) e^{ik\cdot x}\, \nabla _{[3]i} \mathcal{Y}^{{\ell_3}} (S^3).
\end{equation}
Modes containing the $\mathcal{Y}^{{\ell_3} ,\pm} _i $ spherical
harmonics will not mix with other modes because they are in
different representations of $SO(4)$ \cite{spectra}. Hence a third
independent set of  modes is
\begin{equation}\label{type 3}
\textrm{Type 3:} \quad  A_\mu =0,  \quad A_\rho = 0,
 \quad A_i = \phi ^{\pm}_{3} (\rho) e^{ik\cdot x} \,
 \mathcal{Y}^{{\ell_3} ,\pm} _i (S^3).
\end{equation}

There may be other modes with $\partial ^\mu A_\mu \neq 0$.  Such
modes with $k^2=0$ need not be considered because they do not
yield regular solutions.  On the other hand, modes with $\partial
^\mu A_\mu \neq 0$ and $k^2 \not = 0$ can always be put in a gauge
so that they become type 2 modes and hence these modes are
equivalent to modes discussed above in the $\partial^\mu A_\mu =
0$ gauge.

We now compute the spectrum for each type of mode. For type 1 modes,
the equations with $b=\varrho$ and $b=i$ are automatically
satisfied, so we need only consider the equation with $b=\mu$.  This
equation simplifies from (\ref{gaugeEom}) to
\[
\frac{\rp^{7-p}}{(\rho ^2+L^2)^{(7-p)/2}} \partial _\nu
\partial ^\nu A_\mu +\frac{1}{\rho ^3}\partial_\rho (\rho ^3
\partial _\rho A_\mu )+\frac{1}{\rho ^2}\lap3 A_\mu =0 \, ,
\]
and, upon substitution of (\ref{type 1}), becomes
\beq
\frac{1}{\varrho^3}\partial _\varrho ( \varrho^3 \partial_\varrho
 \phi _{1}(\varrho)) +  \frac{\bar{M}^2}{(1+\varrho ^2)^{(7-p)/2}}
  \phi_{1}(\varrho) - \frac{{\ell_3} ({\ell_3} +2)}{\varrho ^2}
   \phi _{1}(\varrho) =0.  \labell{Iphi}
\eeq
The $(p+1)$-dimensional mass spectrum of mesons is then given by
(\ref{mass}) with the dimensionless eigenvalues
$\bm=\bm_1(n,\ell_3)$ determined by this equation.  However,
expanding out the first term in eq.~\reef{Iphi}, one finds that it
precisely matches the previous radial equation \reef{eom2} for the
scalar fluctuations. Hence the spectra of these two sets of
fluctuations are identical, \ie $M_1(n,\ell_3) = M_s(n,\ell_3)$.

For type 2 modes with $\ell_3 = 0$, the gauge fields on the three
sphere vanish, \ie $A_i=0$. The only nontrivial solution of
eq.~\reef{gaugeEom} (setting $b=\mu$) is then $\phi _{2}(\rho) \sim
\rho ^{-3}$. However, requiring a regular solution at the origin
forces us to take the trivial solution $\phi _{2}(\rho) =0 $. Thus
the physical type 2 modes only exist for $\ell_3 \geq 1$.

Putting $b=\mu$ in (\ref{gaugeEom}) and substituting (\ref{type 2})
for the modes, the type 2 radial modes ($\ell_3 \geq 1$) satisfy
\begin{equation}
\frac{1}{\rho} \partial _\rho \left(\rho ^3 \phi _{2}(\rho)\right)
= {\ell_3} ({\ell_3} +2) \tilde{\phi}_{2}(\rho). \labell{IIphis}
\end{equation}
Using (\ref{IIphis}), the equations obtained from (\ref{gaugeEom}) with
$b=\mu$ and $b=\rho$ are equivalent. After using (\ref{mbar}), we
obtain
\beq
\partial _\varrho \left(\frac{1}{\varrho}\, \partial _\varrho
(\varrho ^3 \phi_{2}(\varrho))\right)   -{\ell_3} ({\ell_3} +2)
 \phi_{2}(\varrho) + \frac{\bar{M}^2 \varrho ^2}{(1+\varrho ^2)^{(7-p)/2}}
  \phi _{2}(\varrho)=0, \labell{DpT2}
\eeq
and the solutions of this equation determine the spectrum of meson
masses which we denote as $\bm^2_2(n,\ell_3)$. In this case, if we
expand out eq.~\reef{DpT2} in terms of
$\hat{\phi}(\vr)=\vr\,\phi_2(\vr)$, we find that the result again
matches eq.~\reef{eom2}. Hence the spectrum of the type 2 modes
precisely matches that of the scalar fluctuations and the type 1
gauge fluctuations with $M_2(n,\ell_3) = M_s(n,\ell_3)$ with
$\ell_3 \geq 1$.

Finally, for type 3 modes, the equations (\ref{gaugeEom}) with
$b=\mu$ and $b=\rho$ are automatically satisfied.  The equation
with $b=i$, an $S^3$ coordinate, becomes:
\begin{eqnarray*}
\partial _\mu \partial^\mu A_i + \frac{1}{\rho}\partial _\rho
\left( \frac{\rho (\rho ^2 +L^2)^{(7-p)/2} }{\rp^{7-p}} \partial _\rho A_i
\right) + \frac{(\rho ^2 +L^2) ^{(7-p)/2}}{\rho ^2 \rp^{7-p}}
(\lap3 A_i - R_i ^j A_j) && \\
- \frac{(7-p)}{\rp^{7-p}}(\rho^2 +L^2)^{(5-p)/2} \epsilon_{ijk}
\nabla _{[3]j} A_k &=&0 \, .
\end{eqnarray*}
Substituting (\ref{type 3}) for $A_i$, using the identities
(\ref{harm id}), and making the substitutions (\ref{mbar}), we
find the following equation for $\phi ^\pm _{3} (\vr)$:
\begin{eqnarray*}
\frac{1}{\varrho} \partial _\varrho \left( \varrho
(1+\varrho ^2)^{\frac{7-p}{2}} \partial _\varrho \phi^\pm_{3}
 (\varrho) \right) + \bar{M}^2 \phi _{3} ^\pm(\vr) -
 ({\ell_3} +1)^2 {\frac{(1+\varrho ^2)}{\varrho^2}}^{\frac{7-p}{2}}
  \phi ^\pm_{3} (\varrho) && \\
\mp (7-p)({\ell_3} +1)(1+\varrho ^2)^{\frac{5-p}{2}}
\phi ^\pm _{3}  (\varrho) &=&0 \labell{DpT3}
\end{eqnarray*}
Once again, the spectrum of meson masses is given by (\ref{mass})
with the values of $\bm=\bm_{3,\pm}(n,\ell_3)$ determined from this
equation.  The spectrum of mesons can again be related to the other
spectra with $M_{3 ,\pm} (n,\ell_3) = M_s(n,\ell_3 \pm1)$ with
$\ell_3\ge1$. While in general the spectra can only be evaluated
numerically, the previous relation is established analytically by
mapping eq.~\reef{eom2} to eq.~\reef{DpT3}. One finds that the type
$3,\pm$ modes with $\ell_3=L$ can be related to the solutions
$\phi(\vr)$ of eq.~\reef{eom2} with $\ell_3=L\pm1$ via\footnote{In
more detail, the relation \reef{relate} is established as follows:
Set $\ell_3=L-1$ in eq.~\reef{eom2}. Re-express the equation in
terms of $\phi=\vr^{L-1} F(\vr)$, where the power was chosen to
eliminate the term $(L^2-1)/\vr^2\,\phi$. Next one multiplies the
result by $(1+\vr^2)^{(7-p)/2}$ and takes a derivative with respect
to $\vr$. $F$ only appears in the resulting equation through the
derivatives $F'$, $F''$ and $F'''$ (\ie there are no terms
proportional to $F$) and so the result can be re-expressed as a
second-order equation in terms of $G(\vr)=F'(\vr)/\vr^L$, which one
finds matches precisely eq.~\reef{DpT3} for $\phi^-_{3,\ell_3 =L}$.}
\beqa
\phi^-_{3,\ell_3 =L} &=& \vr^L\partial_\vr \left[\vr^{1-L}
\phi_{\ell_3=L-1} \right]\labell{relate}\\
\phi^{+}_{3,\ell_3 =L} &=& \vr^{-L-2}\partial_\vr \left[\vr^{L+3}
\phi_{\ell_3=L+1} \right]\ . \nonumber\eeqa

There is actually a subtlety for the $\phi^-_3$ profiles. Typically
the radial ODE's yield asymptotic solutions which behave like
$\phi(\vr)=b_1\vr^{-\al_1}+b_2\vr^{-\al_2}$ in the limit
$\vr\rightarrow\infty$. The mass eigenvalues are then determined by
requiring that the coefficient $b_i$ vanishes for the term with
$\al_i$ negative. However, in certain cases with $\phi^-_3$,
eq.~\reef{DpT3} yields solutions where both $\al_i$ are positive.
Specifically for this mode, we have $\al_1=6-p-\ell_3$ and
$\al_2=\ell_3+1$ and so for $\ell_3<6-p$, both modes converge at
infinity. However, in these cases, supersymmetry `selects' the
physical profile as that with $\ell_3+1$, \ie this choice yields a
supersymmetric spectrum.

\subsubsection{Analysis of the spectrum}\label{analysisDpDp4}

In the previous subsection, we computed the spectra of bosonic
mesons in $(p+1)$-dimensional $U(N)$ SYM coupled to a fundamental
hypermultiplet by computing the spectrum of fluctuations of scalar
and gauge fields on the D$(p+4)$-brane worldvolume. Classifying the
massive meson states in representations $(j_1,j_2,j_3)$ of $SU(2)_R
\times SU(2)_L\times SO(4-p)$, where $j_{1,2,3}$ denote the
$SU(2)_{R,L}$ and $SO(4-p)$ spin, the bosonic modes of the
D$(p+4)$-brane give rise to the following mesonic states and mass
spectra:\footnote{Keep in mind that the $SO(4-p)$ group only really
acts for $p\le 2$.}
\begin{itemize}
\item $4-p$ scalars in the $(\frac{\ell_3}{2},\frac{\ell_3}{2},1)$
corresponding to fluctuations transverse to the D$p$-branes, with mass
$M_s(n,\ell_3)$, $n\geq 0, \, \ell_3 \geq 0$;
\item 1 scalar in the $(\frac{\ell_3}{2},\frac{\ell_3}{2},0)$
corresponding to transverse fluctuations but parallel to the
separation of the D$p$- and D($p$+4)-branes, with mass
$M_s(n,\ell_3)$, $n\geq 0, \, \ell_3 \geq 0$;
\item 1 vector in the $(\frac{\ell_3}{2},\frac{\ell_3}{2},0)$
corresponding to type 1 gauge fields on the  D$(p+4)$-brane, with
mass $M_1(n,\ell_3)=M_s(n, \ell_3)$, $n\geq 0, \, \ell_3 \geq 0$;
\item 1 scalar in the $(\frac{\ell_3}{2},\frac{\ell_3}{2},0)$
corresponding to type 2 gauge fields on the  D$(p+4)$-brane, with
mass $M_2(n,\ell_3)=M_s(n, \ell_3)$, $n\geq 0, \, \ell_3 \geq 1$;
\item 1 scalar in the $(\frac{\ell_3-1}{2},\frac{\ell_3+1}{2},0)$
corresponding to type $3,+$ gauge fields on the  D$(p+4)$-brane,
with mass $M_{3,+}(n,\ell_3)=M_s(n, \ell_3+1)$, $n\geq 0, \, \ell_3
\geq 1$;
\item 1 scalar in the $(\frac{\ell_3+1}{2},\frac{\ell_3-1}{2},0)$
corresponding to type $3,-$ gauge fields on the  D$(p+4)$-brane,
with mass $M_{3,-}(n,\ell_3)=M_s(n, \ell_3-1)$, $n\geq 0, \, \ell_3
\geq 1$.
\end{itemize}
In summary, the spectra can be related to each other through
\beq
M_s(n,\ell_3)=M_{1}(n,\ell_3)=M_{2}(n,\ell_3)=M_{3,+}(n,\ell_3-1)
=M_{3,-}(n,\ell_3+1)\ . \labell{D2D6modes}
\eeq
Again, for general $p$, we were unable to find an analytic solution
for the spectrum but the mass eigenvalues are fixed by a simple ODE
\reef{eom2} and may be determined numerically. An analytic solution
was found in \cite{spectra} for p=3 -- see Appendix \ref{solns}.
Another special case is $p=1$ where analytic radial profiles
(written in terms of Bessel functions) can be found for
${\ell_3}=0$. As advertised in the introduction, all of the masses
scale parametrically as $M \propto m_q / \gef(m_q)$, where
$\gef(m_q)$ is the running effective coupling \reef{couple}
evaluated at the quark mass scale.

As noted earlier, the D$p$-D$(p+4)$ brane system preserves eight
supercharges which corresponds to $\mathcal{N}=2$ supersymmetry in
four dimensions.  The mesons, as massive representations of the
supersymmetry algebra, should fill out long supermultiplets. The
construction of these supermultiplets for the $(p+1)$-dimensional
theories with eight supercharges is a simple extension of that for
$\mathcal{N}=2$ multiplets in the four-dimensional theory,
considered in \cite{spectra}. The multiplets are constructed by
acting with the supercharges $Q$ on a state with spin
$\frac{\ell_3}{2}$ under $SU(2)_R$ which is annihilated by the
$\bar{Q}$'s.  Since $SU(2)_L$ commutes with the supercharges, all
states in a given supermultiplet will be in the same representation
of $SU(2)_L$.  Of course, as dictated by supersymmetry, each
multiplet contains an equal number of bosonic and fermionic
components, \ie $8(\ell_3+1)$ of each. For $\ell_3\geq 2$, the
bosonic contributions to the generic multiplet are: $6-p$ real
scalars and one vector in the $\frac{\ell_3}{2}$ of $SU(2)_R$ and
two real scalars in the $\frac{\ell_3}{2} \pm 1$ of $SU(2)_R$. Now
the fermions can be represented as Dirac spinors with two ($p=1,2$)
or four ($p=3,4$) components. Hence, for $p=3,4$, the generic
multiplet contains a single Dirac fermion, in each of
${\ell_3+1\over2}$ and ${\ell_3-1\over2}$ of $SU(2)_R$. For $p=1,2$,
there are two Dirac fermions in each of the $\frac{\ell_3+1}{2}$ and
$\frac{\ell_3-1}{2}$ . However, in the latter case, there is also a
nontrivial $SO(4-p)$ action and these pairs of spinors combine
together in the spin-half representation of this rotation
group.\footnote{These considerations may be extended to $p=0$, where
the dual theory is supersymmetric matrix quantum mechanics coupled
to degrees of freedom in the fundamental representation. The latter
include four fermionic variables, which are organized as a spinor of
the $SO(4)$ symmetry.} Exceptional semi-short supermultiplets appear
for $\ell_3=0,1$. For $\ell_3=0$, the spectrum contains $5-p$
scalars and one vector which are singlets of $SU(2)_R$, one scalar
in the 1, and one ($p$=3,4) or two ($p$=1,2) Dirac fermions in the
$1\over2$. For $\ell_3=1$, there are $6-p$ scalars and one vector in
the ${1\over2}$, one scalar in the ${3\over2}$ and one ($p$=3,4) or
two ($p$=1,2) Dirac fermions in each of the 0 and 1.

In each of the cases above, the bosonic content is, of course,
identical to that found using supergravity. Hence supersymmetry
allows us to extend the computed bosonic spectrum to include
fermions. First we have for $p=1,2$
\begin{itemize}
\item 1 fermion in the $(\frac{\ell_3+1}{2},\frac{\ell_3}{2},\frac{1}{2})$
with mass $M_{f,1}(n,\ell_3)=M_s(n, \ell_3)$, $n\geq 0, \, \ell_3 \geq 0$;
\item 1 fermion in the $(\frac{\ell_3}{2},\frac{\ell_3-1}{2},\frac{1}{2})$
with mass $M_{f,2}(n,\ell_3)=M_s(n, \ell_3+1)$, $n\geq 0, \, \ell_3 \geq 0$.
\end{itemize}
Keeping in mind that the spinor representations are twice as large
for $p=3,4$, we can keep the classification as above except for
dropping $j_3$ as there is no $SO(4-p)$.


\subsubsection{Meson spectrum for $p=2$}\label{xam}

As a specific example, let us consider the case $p=2$ in which the
background geometry is induced by $N$ coincident D2-branes and
there is one D6-brane probe embedded a distance $L$ in the
789-directions:
\begin{equation}
\begin{array}{ccccccccccc}
& 0 & 1 & 2 & 3 & 4 & 5& 6&7&8&9\\
D2 & \x & \x & \x &   &   & &  & & & \\
D6 & \x & \x & \x & \x& \x  & \x & \x &  &  &  \\
\end{array}
\labell{D2-D6}\end{equation}
The radial differential equation \reef{eom2} (for scalar fluctuations of the D6-brane) is
\beq
\partial_\rho^2 \phi(\vr)+\frac{3}{\vr}\partial_\vr \phi(\vr) +
\left(
\frac{\bm^2}{(1+\vr^2)^{5/2}}-\frac{{\ell_3}({\ell_3}+2)}{\vr^2}\right)
\phi (\vr) =0 \labell{radialDE}\eeq
which can be solved using the shooting method.  For either $\vr
\to 0$ or $\vr \to \infty$, the solution has the form $\phi (\vr)
= A \vr ^{\ell_3} + B \vr ^{-{\ell_3}-2}$.  Using the boundary condition
$\phi(\vr \to 0) = \vr^{\ell_3}$, we solved (\ref{radialDE})
numerically.  As the solutions to (\ref{radialDE}) must be regular
for all values of $\vr$, we tuned the constant $\bm$ to obtain the
regular solutions, which behave as $\phi = \vr ^{-{\ell_3} -2}$ for
$\vr \to \infty$.  In this way, the three-dimensional mass
spectrum of scalar mesons was found to be
\beq
M^2 = \frac{4\pi}{3}\bm^2\frac{m_q^3}{\gym^2 N} \, ,
\labell{d2exam}
\eeq
where the values of $\bm_s(n,\ell_3)$ are given in table
\ref{D2D6spectrum}.  As noted earlier, the eigenvalues for the type
1 and 2 gauge fields are identical to those for the scalar modes and
hence their spectrum is also given by \reef{d2exam}.  For the type 3
gauge field modes, we solved \reef{DpT3} using the shooting method
and found eigenvalues identical to those in table \ref{D2D6spectrum}
but with the $\ell_3$ label shifted as indicated in
eq.~\reef{D2D6modes}, \ie $M_{3 ,\pm} (n,\ell_3) = M_s(n,\ell_3
\pm1)$, $\ell_3 \ge 1$.

\TABLE{
\begin{tabular}{|c|c|c|c|c|c|c|c|}
\cline{3-8}
\multicolumn{2}{c}{ } & \multicolumn{6}{|c|}{$n$} \\ \cline{3-8}
\multicolumn{2}{c|}{ } & 0 & 1 & 2 & 3 & 4 & 5 \\ \hline
 &0    & 11.34 & 36.53 & 75.49 & 128.19 & 194.65 & 274.86 \\
\cline{2-8}
$\ell_3 $ & 1    & 33.39 & 70.37 & 121.01 & 185.36 & 263.41 & 355.20\\
\cline{2-8}
       &  2  & 66.20 & 114.96 & 177.33 & 253.36 & 343.08 & 446.49\\
\cline{2-8}
       &  3      & 109.75 & 170.30 & 244.43 & 332.17 & 433.57 & 548.65\\
\hline
\end{tabular}
\caption{The mesonic spectrum in terms of $\bar{M}_s^2(n,\ell_3)$
corresponding to a configuration a D6-brane probe in the D2-brane
geometry. Each mass eigenvalue is labelled by the quantum numbers
$\ell_3$ and $n$ (where $n$ represents the number of nodes in the
$\phi(\rho)$ function).} \label{D2D6spectrum} }


\subsection{Meson spectrum in D$p$-D$(p+2)$}

Consider the following configuration of $N$ coincident D$p$-branes
($1\leq p < 5$) and one D$(p+2)$-brane probe, embedded a distance
$L$ from the D$p$-branes:
\begin{equation}
\begin{array}{cccccccccccc} & 0 & 1 & \cdots & p-1 & p& p+1
& p+2 & p+3 &p+4 & \cdots & 9\\
Dp & \x & \x & \cdots &  \x & \x  & &   &  &  &  &  \\
D(p+2) & \x & \x & \cdots & \x &  & \x & \x & \x & & &  \\
\end{array}\labell{Dp-D(p+2)}
\end{equation}
As before, this orientation of branes was chosen to produce a
supersymmetric system which is in static equilibrium.  Inserting the
D$(p+2)$-brane breaks the $SO(1,p)$ symmetry of the $01\cdots
p$-directions to $SO(1,p-1)$ acting on $01\cdots (p-1)$. Similarly
the $SO(9-p)$ symmetry of the space transverse to the D$p$-branes to
$SO(3)=SU(2)_R$ acting on $Y^1Y^2Y^3$ and, for $L\ne 0$, $SO(5-p)$
acting on $Y^4\cdots Y^{8-p}$. The R-symmetry group for fields on
the D$(p+2)$-brane is $SU(2)_R$. The induced metric on the
D$(p+2)$-brane is
\beq ds^2 = \left(\frac{\sqrt{\rho ^2
+L^2}}{\rp}\right)^{\frac{7-p}{2}}ds^2(\mathbb{E}^{(1,p-1)}) +
\left(\frac{\rp}{\sqrt{\rho ^2 +L^2}}\right)^{\frac{7-p}{2}}(d\rho
^2 +\rho^2 d\Omega_2^2) \labell{gindDp2} \eeq
where, as usual,  we have introduced spherical polar coordinates
(with radial coordinate $\rho$) in the probe brane worldvolume
directions orthogonal to the background D$p$-branes.

The dual gauge theory is $(p+1)$-dimensional, but the fundamental
hypermultiplet has been introduced on a $p$-dimensional surface.
Hence the matter fields are localized on a codimension-one defect
\cite{defect}, \eg $x^p=0$.  The theory again has eight conserved
supercharges and the quark mass is still given by (\ref{qmass}).
Together the rotation symmetries above, \ie $SU(2)_R\times SO(5-p)$,
form the R-symmetry group of the gauge theory.

We follow the same procedure described for the D$p$-D$(p+4)$ system
to compute the spectrum of mesons corresponding to fluctuations of
the D$(p+2)$-brane.   In this case, however, the D$(p+2)$-brane can
fluctuate in the $X^p$-direction, parallel to the background branes,
as well as in the $Y^4\cdots Y^{9-p}$-directions, transverse to the
background D$p$-branes. We take $\chi^A$ ($A=4,...,9-p$) and $\psi$
as the fluctuations in directions transverse and parallel to the
background D$p$-branes, respectively:
\beqa
Y^A &=& \delta ^A _9 \,L + 2 \pi \al' \chi^A, \quad A = 4, ..., 9-p \\
X^p &=& 2\pi\alpha ' \psi.
\eeqa

As with the D$p$-D$(p+4)$-brane configuration, the relevant action
for the D$(p+2)$-brane fields is the DBI action \reef{DBI} combined
with the Wess-Zumino involving $C_{(p+1)}$:
\beq S = S_{DBI} - 2 \pi \al' \tau_{p+2} \int P[C_{(p+1)}]\wedge F
\, . \labell{actionDpDp2} \eeq
In this case, the Wess-Zumino term produces a coupling between the
gauge fields and the scalar $\psi$.

Considering first the fluctuations $\chi^A$ in directions orthogonal
to the background branes, the relevant quadratic Lagrangian density
is
\beq  \mathcal{L} = -\frac{(2\pi \al')^2}{2} \tau _{p+2}
 e^{-\phi} \sqrt{-\det g_{ab}}  \,  \left(\frac{r_p}{r}\right)^{\frac{7-p}{2}} g^{cd}
\partial_c \chi ^A \partial_d \chi^A
\labell{LagP+2}\eeq
where summation over $A = 4,...,9-p$ is implied. Here we have
dropped the `constant' term proportional to the determinant of the
induced metric $g_{ab}$ and the dilaton, since it is again
independent of the fluctuations. Now in an expansion to quadratic
order the entire pre-factor in eq.~\reef{LagP+2} can be treated as
though it is independent of $\chi^A$. In particular, we can take
$g_{ab}$ to be given by (\ref{gindDp2}).

As each of the $\chi^A$ fluctuations appears on equal footing in the
Lagrangian (\ref{LagP+2}), the superscript is dropped in the
following. Now we expand the modes as
\beq \chi = \phi_\perp (\rho)\, e^{ik\cdot x}\,\mathcal{Y}^{\ell_2}
(S^2) \, , \labell{ModeExpP+2} \eeq
where $e^{ik\cdot x}$ is a plane wave in the $0,1,...,p-1$ space and
$\mathcal{Y}^{\ell_2} (S^2)$ are spherical harmonics on an $S^2$ of
unit radius ($\ell_2 = 0,1,2,...$).  With $\varrho$ and $\bm$ as
defined in (\ref{mbar}), the equation determining the radial profile
is
\beq
\partial_\vr^2 \phi_\perp(\vr) +\frac{2}{\vr}\partial_\vr \phi_\perp(\vr) +
\left( \frac{\bm^2}{(1+\vr^2)^{(7-p)/2}}-\frac{\ell_2
(\ell_2+1)}{\vr^2}\right)
\phi_\perp(\vr) =0. \labell{transp2} \eeq
As discussed in section \ref{mesonsDpDp4}, the solutions
corresponding to physical mesons are real, regular at the origin and
convergent asymptotically. Analytic solutions to this equation for
$p=3$ (in terms of hypergeometric functions) are given in appendix
\ref{solns}.\footnote{Eq.~\reef{transp2} also becomes a
hypergeometric equation for $p=5$ but there are no solutions
satisfying all of the necessary criteria.} For $p=1,2,4$, we solved
\reef{transp2} numerically (using the shooting method) to determine
the mass eigenvalues.  As in the previous section then, for $1 \leq
p \leq 4$, the spectrum of mesons (here confined to a
($p$--1)-dimensonal defect) is given by (\ref{mass}) with the
dimensionless eigenvalues $\bm=\bm_\perp(n,\ell_2)$ computed from
\reef{transp2}, where $n$ is the number of nodes of the
$\phi_\perp(\rho)$ function.

The relevant Lagrangian density for fluctuations of the gauge fields
and $\psi$, the fluctuation of the D$(p+2)$-brane along the $X^p$
direction, follows from \reef{actionDpDp2} as
\beqa \mathcal{L} = -(2 \pi \al') ^2 \tau_{p+2} \left[ e^{-\phi}
\sqrt{-\det g_{ab}} \right. &&  \left( \frac{1}{2}
\left(\frac{r}{r_p}\right)^{\frac{7-p}{2}}
 \partial_c \psi \partial_d \psi   + \frac{1}{4}F_{cd}F^{cd} \right)   \nonumber \\
&&\left. - \left(\frac{r}{r_p}\right)^{7-p} \left( F_{\theta
\varphi}
\partial_\rho \psi +   F_{\varphi \rho} \partial_\theta \psi +
F_{\rho \theta} \partial _\varphi \psi \right) \right], \eeqa
where $\rho, \theta, \varphi$ are spherical polar coordinates in the
$Y^1Y^2Y^3$-space. To quadratic order, the equations of motion are
\beqa
\partial_a \left[ \rho ^2 \sqrt{\det h} \left(\frac{r}{r_p}
\right)^{\frac{7-p}{2}} g^{ab} \partial_b \psi \right]
-(7-p) \frac{\rho r^{5-p}}{r_p^{7-p}} \epsilon ^{ij}\nabla_{[2]i}A_j &=&0  \labell{Dp2EOM1} \\
\partial_a \left( \rho ^2 \sqrt{\det h} F^{ab} \right)
-(7-p) \rho \frac{ r^{5-p}}{r_p^{7-p}} \epsilon ^{bj}\partial_i \psi &=&0 \labell{Dp2EOM2}
\eeqa
where we are using the index notation given in footnote
\ref{indexConv}.  Also $h_{ij}$ and $\epsilon^{ij}$ are the metric
and antisymmetric tensor density on the unit two-sphere,
respectively. Note that $\epsilon ^{\theta \varphi} = 1$. The second
term in each equation results from the Wess-Zumino term and in
\reef{Dp2EOM2} this term is only present if $b$ is an index on the
$S^2$.

We can expand the scalar and gauge fields in terms of spherical
harmonics on the two-sphere component of the D$(p+2)$-brane: $A_\mu,
A_\rho, \psi$ in terms of scalar spherical harmonics and $A_i$, in
terms of vector spherical harmonics. Note that the equations of
motion \reef{Dp2EOM1} and \reef{Dp2EOM2} imply that the scalar field
$\psi$ couples only to the $A_i$ gauge field modes. Thus, working in
the gauge $\partial^\mu A_\mu =0$, we can define one type of mode
not coupled to $\psi$ as
\beq
\textrm{Type 1:}\quad  A_\mu = \xi_\mu  \phi_1 (\rho) e^{ik\cdot x}
\mathcal{Y}^{\ell_2}(S_2) \, , \quad \xi\cdot k =0 \,
,\quad A_\rho =0 \,, \quad A_i =0 \,. \labell{type1pp2}
\eeq
For $\psi =0$, there is a second type of mode given by
\beq
\textrm{Type 2:}\quad  A_\mu = 0 \, , \quad A_\rho =\phi_2(\rho)
e^{ik\cdot x}\mathcal{Y}^{\ell_2}(S_2)  \,, \quad A_i =
\tilde{\phi}_2(\rho)e^{ik\cdot x}\nabla_{[2]i} \mathcal{Y}^{\ell_2}(S_2)
\,. \labell{type2pp2}
\eeq
Finally, there is a third type of gauge field mode which is coupled
to $\psi$ for $\ell_2 \ge 1$:
\beq \textrm{Type 3:} \quad  A_\mu = 0 \, , \quad A_\rho =0 \, ,
\quad A_i = \phi_{3}(\rho ) e^{ik\cdot x}  \sqrt{\det
h}\epsilon_{ij} \nabla_{[2]}^i \mathcal{Y}^{\ell_2}(S^2) \, ,
\labell{type3pp2} \eeq
where the factor $\sqrt{\det h}$ makes up for the density weight of
the two-dimensional $\epsilon$-symbol.

We now proceed to compute the spectra for each type of mode. For
type 1 modes, the equation of motion is \reef{eom2} with $b=\mu$.
Substituting \reef{type1pp2} and making the redefinitions
\reef{mbar}, we obtain
\beq
\partial _\varrho ^2 \phi_1(\varrho) + \frac{2}{\vr} \partial_\vr
\phi_1(\varrho) + \left(\frac{\bm^2}{(1+\vr^2)^{\frac{7-p}{2}}} -
\frac{\ell_2 (\ell_2 +1)}{\vr^2} \right)  \phi_1(\varrho) =0. \labell{gaugeT1p2}
\eeq
Note that this result is identical to eq.~\reef{transp2}, the
differential equation for the scalar fluctuations transverse to the
background branes. Thus, the mass spectrum $\bm=\bm_1(n,\ell_2)$
here will be identical to that for the transverse scalars.

For the type 2 gauge fields and $\psi =0$, \reef{Dp2EOM1} is
identically satisfied.  Note that for $\ell_2 =0$, $A_i=0$.  Then
\reef{Dp2EOM2} with $b=\mu$ yields $\phi_2 \sim \rho^{-2}$.  As the
latter is not regular at the origin, the only solution for $\ell_2
=0$ is trivial (\ie $\phi_2 =0$) and we need only consider $\ell_2
\geq 1$ for these modes.

For $\ell_2 \geq 1$, \reef{Dp2EOM2} with $b=\mu$ gives
$\partial_\rho (\rho ^2 \phi_2)=\ell_2 (\ell_2+1) \tilde{\phi}_2$.
With this, the equations obtained from \reef{Dp2EOM2} with $b=\rho$
and $b=i$ are equivalent and give, with the definitions as in
eq.~\reef{mbar},
\beq
\partial_\vr ^2 \phi_2(\vr) + \frac{4}{\vr} \partial_\vr \phi_2(\vr) +
\left( \frac{\bm^2 }{(1+\vr^2)^{\frac{7-p}{2}}} +
\frac{2-\ell_2(\ell_2+1)}{\vr^2}\right) \phi_2(\vr) =0 \, ,
\labell{gaugeT2p2}
\eeq
which defines the mass spectrum which we denote as $\bm^2_2(n,
\ell_2)$, $\ell_2 \geq 1$. Putting $\hat{\phi}_2 = \vr \phi_2$, this
equation becomes the ODE \reef{transp2} and thus the spectrum is
again identical to that for the transverse scalars and type 1 gauge
fields: $M_2(n, \ell_2)=M_\perp(n, \ell_2) = M_1(n, \ell_2)$.

As noted above, the type 3 gauge field modes are coupled to the
scalar field $\psi$ which represents fluctuations of the probe brane
along the $X^p$ direction, parallel to the background D$p$-branes.
The mode $\ell_2=0$ is an exception since the gauge field vanishes:
$A_i=0$.  Then, with $\psi = e^{ik\cdot x} \phi_{||} (\rho )$ and
the redefinitions \reef{mbar}, eq.~\reef{Dp2EOM1} yields
\beq
\partial^2_\vr \phi_{||}(\vr) + \left(\frac{2}{\vr}+\frac{(7-p)\vr}{1+\vr^2}
 \right) \partial_\vr \phi_{||}(\vr) + \frac{\bm^2 }{(1+\vr^2)^{\frac{7-p}{2}}}
    \phi_{||}(\vr) =0. \labell{parSclr}
\eeq
Solving this equation and imposing regularity requirements yields
the spectrum of mesons $M_{||}(n, 0)=M_{\perp}(n,1)$.

For $\ell_2 \geq 1$, we proceed via separation of variables,
expanding the scalar field as $\psi = e^{ik\cdot x} \phi_{||} (\rho
) \mathcal{Y}^{\ell_2}(S^2)$ and using \reef{type3pp2} for the gauge
field, so that \reef{Dp2EOM1} becomes
\beq
\left(\frac{r_p}{r} \right)^{7-p} M^2  \phi_{||}(\rho) +
\frac{1}{\rho^2 r^{7-p}} \partial_\rho \left( \rho^2 r^{7-p}
\partial_\rho \phi_{||} (\rho) \right) -\frac{\ell_2 (\ell_2 +1)}{\rho^2 }
\left( \phi_{||}(\rho) + (7-p)\frac{\rho}{r^2} \phi_{3}(\rho)\right)  =0 \, ,
\labell{coupled1}
\eeq
while \reef{Dp2EOM2} with $b=i$ gives
\beq
\left(\frac{r_p}{r} \right)^{7-p} M^2 \phi_{3}(\rho) + \frac{1}{r^{7-p}}
\partial_\rho \left( r^{7-p} \partial_\rho  \phi_{3}(\rho) \right)
-\frac{\ell_2 (\ell_2 +1)}{\rho^2 }\left(  \phi_{3}(\rho) +
(7-p)\frac{\rho^3}{r^2} \phi_{||} (\rho)\right)  =0\ .
\labell{coupled2} \eeq
In both of these equations, $r =\sqrt{\rho^2 +L^2}$.  We diagonalize
this system of equations by defining two new radial functions
$\tilde{\phi} _+(\rho)\,, \, \, \tilde{\phi} _-(\rho)$ as
\beq
\tilde{\phi} _+(\rho) = {\ell_2} \phi_{3}(\rho) +{\rho}\phi_{||}(\rho)\, , \quad
\tilde{\phi} _-(\rho) = (\ell_2+1)\phi_{3}(\rho) -{\rho}\phi_{||}(\rho)\,  . \labell{diagFields}
\eeq
With these new functions and also the definitions \reef{mbar},
\reef{coupled1} and \reef{coupled2}, the decoupled equations become
\beq
\begin{array}{r}
\partial^2_\vr \tilde{\phi} _+(\vr) +\frac{(7-p)\vr}{1+\vr^2}
\partial_\vr \tilde{\phi} _+(\vr)
+ \left( \frac{\bm^2 }{(1+\vr^2)^{\frac{7-p}{2}}} -
\frac{\ell_2(\ell_2+1)}{\vr^2}
-(7-p)\frac{(\ell_2+1)}{1+\vr^2} \right) \tilde{\phi} _+(\vr) = 0 \, \, \\
\partial^2_\vr \tilde{\phi} _-(\vr) +\frac{(7-p)\vr}{1+\vr^2} \partial_\vr \tilde{\phi} _-(\vr)
+ \left( \frac{\bm^2 }{(1+\vr^2)^{\frac{7-p}{2}}} -
\frac{\ell_2(\ell_2+1)}{\vr^2} +(7-p)\frac{\ell_2}{1+\vr^2} \right)
\tilde{\phi} _-(\vr) = 0\, .
\end{array} \labell{coupledEOM}
\eeq
As usual, by solving these equations and imposing regularity
requirements, the eigenvalues $\bm$ are found for $1 \leq p \leq 4$
(while for $p=5$ there are no normalizable modes).\footnote{Here
again, one finds a subtlety for the $\tilde{\phi}_-$ profiles. In
the limit $\vr\rightarrow\infty$, the solutions behave like
$\tilde{\phi}(\vr)=b_1\vr^{-\al_1}+b_2\vr^{-\al_2}$ where
$\al_1=6-p-\ell_2$ and $\al_2=\ell_2$. Hence both $\al_i$ are
positive for $\ell_2<6-p$ and both modes converge at infinity.
Hence, in these cases, we still determine the physical masses by
demanding that $b_1$ vanish, which yields a supersymmetric
spectrum.} We denote the eigenvalues for $\tilde{\phi}_+(\rho)$ and
$\tilde{\phi}_-(\rho)$ respectively as $\bm_+(n,\ell_2)$ and
$\bm_-(n,\ell_2)$.  The spectrum of mesons corresponding to
fluctuations of the D$(p+2)$-brane in the $p$-direction and
fluctuations of the gauge fields is then given by \reef{mass} with
these values of $\bm^2$ and these can be related to the spectrum of
scalar fluctuations transverse to the background branes via
$M_{\pm}(n,\ell_2) = M_\perp (n, \ell_2 \pm 1)$, $\ell_2 \geq 1$.
Again, the spectra are only evaluated numerically in general,
however, this matching of the spectra is established analytically by
mapping eq.~\reef{transp2} to eq.~\reef{coupledEOM}. One finds that
the $\tilde{\phi}_\pm$ modes are related to the $\phi_\perp$ modes
as follows:
\beqa
\tilde{\phi}_{-,\ell_2 =L} &=& \vr^L\partial_\vr
\left[\vr^{1-L} \phi_{\perp,\ell_2=L-1} \right]\labell{relatedx2}\\
\tilde{\phi}_{+,\ell_2 =L} &=& \vr^{-L-1}\partial_\vr
\left[\vr^{L+2} \phi_{\perp,\ell_2=L+1} \right]\ .\nonumber \eeqa
%


In computing the spectra of fluctuations of scalar and gauge fields
on the D$(p+2)$-brane worldvolume, we have found the spectra of
mesons living on a codimension-one defect in the ($p+1$)-dimensional
super-Yang-Mills theory.  From the supergravity computations, there
are $6-p$ scalar mesons corresponding to transverse fluctuations of
the D$p$-branes with mass $M_\perp$, one vector meson corresponding
to type 1 gauge fields with mass $M_1$, one scalar meson
corresponding to type 2 gauge fields with mass $M_2$, and two
scalars, corresponding to the $\tilde{\phi}_\pm$ modes with masses
$M_{\pm}$.  The spectra for these different modes are simply related
with
\beq
\bm^2 _\perp (n,\ell_2) = \bm^2 _1 (n,\ell_2) =\bm^2 _2 (n,\ell_2)=
 \bm^2 _+ (n,\ell_2-1) = \bm^2 _- (n,\ell_2+1) \, . \labell{DpDp2modes}
\eeq
As with the D$p$-D$(p+4)$ brane configuration, we were unable to find
an analytic solution for the spectrum for general $p$.  However, the
mass eigenvalues are fixed by the ODE \reef{transp2} and can be
easily computed using numerical techniques.  The notable exception
is the case $p=3$ for which solutions were found in terms of
hypergeometric functions -- see Appendix \ref{solns}.  Hence while
the precise numbers change, the masses scale as $M \propto m_q /
\gef(m_q)$, which is {\it identical} to the scaling found in the
D$p$-D$(p+4)$ brane system.

As discussed in section \ref{analysisDpDp4}, the mesons should fill
massive supermultiplets.  Counting the bosonic contributions found
using supergravity, we see that each multiplet contains eight
bosonic degrees of freedom, as expected from the dual gauge theory.


\subsection{Meson spectrum in D$p$-D$p$}\label{noway}

Consider the following configuration of $N$ coincident D$p$-branes
($2\leq p< 5$) and one D$p$-brane probe, in which the probe brane is
a distance $L$ from the background branes:
\begin{equation}
\begin{array}{cccccccccccc}
& 0 & 1 & \cdots & p-2& p-1 & p& p+1 & p+2 & p+3  & \cdots & 9\\
{\rm background} & \x & \x & \cdots &  \x & \x & \x  & &   &  & & \\
{\rm probe} & \x & \x & \cdots & \x& &   & \x & \x &  & &   \\
\end{array}\labell{Dp-Dp}
\end{equation}
The orientation of D-branes was again chosen to preserve
supersymmetry and the branes are in static equilibrium.  Embedding
the probe brane in this way reduces the number of supercharges from
sixteen to eight.  Correspondingly, the isometry groups of the
background geometry \reef{nearGeom} are broken.  The $SO(1,p)$
symmetry of the $01\cdots p$-directions has been reduced to
$SO(1,p-2) \times SO(2)$. Further the $SO(9-p)$ symmetry
corresponding to rotations in the $Y^1\cdots Y^{9-p}$-directions is
broken to $SO(2)'$ acting in the $(p+1,p+2)$-plane and for $L\ne 0$,
$SO(6-p)$ acting in the remaining transverse directions orthogonal
to the separation of the branes.  The induced metric $g_{ab}$ on the
probe brane is
\beq
ds^2 = \left( \frac{\sqrt{\rho^2 +L^2}}{r_p}\right)^{\frac{7-p}{2}}ds^2
(\mathbb{E}^{(1,p-2)})+ \left( \frac{r_p}{\sqrt{\rho^2 +L^2}}\right)^{\frac{7-p}{2}}
\left( d\rho^2 +\rho ^2 d\theta^2 \right) \, ,
\eeq
where, as usual, we are using polar coordinates $\rho,\theta$ in the
probe brane worldvolume directions transverse to the background
D$p$-branes.

The dual description has $(p+1)$-dimensional super-Yang-Mills
coupled to a fundamental hypermultiplet confined to a
$(p-1)$-dimensional surface. That is, the matter fields live on a
codimension-two defect \cite{defect}. The quark mass \reef{qmass}
again corresponds to the mass of a fundamental string stretching
between the background branes and the probe brane. The theory has
eight conserved supercharges and the R-symmetry group has two
components: the $SO(6-p)$ rotations and the diagonal rotations in
$SO(2)\times SO(2)'$.

For this brane configuration, the fluctuations of the probe brane
fall into two classes: those orthogonal and parallel to the
background branes. We write the fluctuations around the fiducial
embedding as
\beqa
Y^A &=& \delta^A _9 L + 2 \pi \al' \chi^A, \quad A= 3, ..., 9-p \, ;\\
X^B &=& 2 \pi \al' \psi^B, \quad B = p-1,p \, .
 \eeqa
The action defining the dynamics of the D$p$-brane probe is given by
the DBI action \reef{DBI} plus the Wess-Zumino term with
$C_{(p+1)}$:
\beq S=S_{DBI} -  \tau_p \int P[C_{(p+1)}] \, . \labell{actionDpDp}
\eeq

For the `orthogonal' fluctuations, the quadratic Lagrangian density
is
\beq \mathcal{L} = - \frac{(2\pi \al')^2}{2} \tau _{p} e^{-\phi}
\sqrt{-\det g_{ab}} \left(\frac{r_p}{r}\right)^{\frac{7-p}{2}}g^{cd}
\partial_c \chi ^A \partial_d \chi^A
 \labell{lagP}
\eeq
where summation over $A$ is implied and $r^2 = L^2 + \rho^2$.  As
before, the quadratic Lagrangian depends only on the derivatives of
$\chi^A$. Taking $\chi$ to be any one of the $\chi^A$, we expand the
modes as
\beq \chi = \phi _\perp (\rho)\, e^{ik\cdot x} e^{i\ell \theta} \, ,
\labell{ModeExpP} \eeq
where $e^{ik\cdot x}$ is a plane wave in the $(0,1,...,p-2)$-space
and $\ell=0,\pm 1, \pm 2 ,...$.  The equation of motion reduces to
determining the radial profile $\phi_\perp(\vr)$:
\beq
\partial_\vr^2 \phi_\perp(\vr) +\frac{1}{\vr}\partial_\vr \phi_\perp(\vr) +
\left( \frac{\bm^2}{(1+\vr^2)^{(7-p)/2}}-\frac{\ell^2}{\vr^2}\right) \phi_\perp(\vr) =0 \,
\labell{transp}
\eeq
where we use the definitions (\ref{mbar}). The solutions with
$\ell=0$ are not normalizable and so we restrict to $\ell$ nonzero.
Since the equation above is symmetric in $\ell$, modes with the same
absolute value of $\ell$ have the same mass. For $p=3$ analytic
solutions in terms of hypergeometric functions are given in appendix
\ref{solns} while for $p=2,4$ solutions were found
numerically.\footnote{Here and for the remaining modes, one finds
there are no normalizable modes for $p=5$.} Thus, for $2\leq p \leq
4$, the spectrum of mesons living on the $(p-1)$-dimensional defect
is given by (\ref{mass}) with $\bm=\bm_\perp(n,|\ell|)$ ($\ell \ne
0$) computed from eq.~\reef{transp}.

We now turn to the `parallel' fluctuations $\psi^{p-1},\psi^p$. In
this case, the Wess-Zumino term introduces a coupling
between these two fields. The quadratic Lagrangian density following
from eq.~\reef{actionDpDp} is
\beq \mathcal{L} = -\frac{(2 \pi \al')^2}{2} \tau_p \left[\rho
\left( \frac{r}{r_p}\right)^{\frac{7-p}{2}} g^{cd}
\partial_c \psi^B \partial_d \psi^B -\left( \frac{r}{r_p}\right)^{7-p}
(\partial_\rho \psi^{p-1}\partial_\theta \psi^p -\partial_\rho \psi^{p}
\partial_\theta \psi^{p-1} ) \right] \,\labell{quadxx}
\eeq
where $r^2 = L^2 +\rho^2$ again. The $\psi^{p-1}, \psi^p$ mixing can
be diagonalized by working with the field $\Psi = \psi^{p-1} + i
\psi^p$ and its complex conjugate $\Psi^\ast  = \psi^{p-1} - i
\psi^p$, in terms of which eq.~\reef{quadxx} becomes
\beq \mathcal{L} = (\pi \al')^2 \tau_p \left[-2\rho \left(
\frac{r}{r_p}\right)^{\frac{7-p}{2}} g^{cd}\partial_c \Psi
\partial_d \Psi^\ast  -i \left( \frac{r}{r_p}\right)^{7-p}
(\partial_\rho \Psi^\ast \partial_\theta \Psi -\partial_\rho \Psi
\partial_\theta \Psi^\ast) \right] \, , \eeq
and the resulting equation of motion is
\beq
\partial_a \left[\rho  \left( \frac{r}{r_p}\right)^{\frac{7-p}{2}}
g^{ab}\partial_b \Psi  \right] -\frac{(7-p)}{2i}\rho
\frac{r^{5-p}}{r_p^{7-p}} \partial_\theta \Psi =0 \, .\labell{heom}
\eeq
Proceeding via separation of variables, we expand the field as $\Psi
= \phi_{||}(\rho)\, e^{ik\cdot x} e^{i\ell \theta}$ and
eq.~\reef{heom} reduces to
\beq
\partial^2 _\vr \phi_{||}(\vr) + \left(\frac{1}{\vr}+
\frac{(7-p)\vr}{1+\vr^2} \right) \partial_\vr \phi_{||}(\vr) +
\left(\frac{\bm^2}{(1+\vr^2)^{\frac{7-p}{2}}} -
\frac{\ell^2}{\vr^2} -\frac{(7-p)\ell}{1+\vr^2} \right)
\phi_{||}(\vr) =0 \, , \labell{scalarParP} \eeq
where we have substituted in the definitions \reef{mbar}.  As
before, we found analytic solutions for $p=3$ (see appendix
\ref{solns}) and numerical solutions for $p=2,4$.

As seen in previous sections, there is a mapping from
eq.~\reef{transp} to eq.~\reef{scalarParP}. One finds that the
$\phi_{||}$ modes can be related to the `orthogonal' modes
via\footnote{Note that is a subtlety here in that there is no
mapping for $L=-1$! Hence supersymmetry seems to dictate that the
$\ell =-1$ solutions of eq.~\reef{scalarParP} are unphysical.
\label{lminus1}}
\beqa
\phi_{||,\ell=L} &=& \vr^{-L-1} \partial_\vr\left[\vr ^{L+1}
\phi_{\perp,|\ell|=L+1} \right], \qquad L\geq 0\labell{relatedx3}\\
\phi_{||,\ell=-L} &=& \vr^{L-1} \partial_\vr\left[\vr ^{1-L}
\phi_{\perp,|\ell|=L-1} \right], \qquad L\geq 2\ .\nonumber \eeqa
Hence the spectra can be matched analytically, even if they are only
evaluated numerically. We denote the spectra $\bm^2_{\pm}(n,|\ell|)$
for $\ell \geq0$ and $\ell\leq-2$, respectively. Thus it follows
from eq.~\reef{relatedx3} that
$\bm_{+}(n,\ell)=\bm_\perp(n,|\ell|+1)$ and $\bm _{-}(n, \ell) =
\bm_\perp (n,|\ell|-1)$.

The spectrum for $\Psi^\ast$ is obtained by noting that the radial
ODE for this field is identical to eq.~\reef{scalarParP} with the
replacement $\ell \to -\ell$. Hence for $\Psi^\ast$, we have:
$\bm_{\ast ,+}(n,\ell)=\bm _{-}(n, -\ell) = \bm_\perp
(n,|\ell|-1)$ with $\ell \ge 2$; and $\bm _{\ast,-}(n,
\ell)=\bm_{+}(n,-\ell)=\bm_\perp(n,|\ell|+1)$ with $\ell \leq
0$.

Finally, we turn to computing the spectrum for gauge field
fluctuations.  The linearized equation of motion comes entirely from
the DBI part of the action \reef{actionDpDp} and is
\beq
\partial_a (e^{-\phi} \sqrt{-\det g_{cd}} F^{ab} ) =
\partial_a (\rho F^{ab}) =0 \, . \labell{DpGaugeEOM}
\eeq
Following the earlier analysis, we find that there are two types of
modes:
\beqa
 \textrm{Type 1:}&& \quad  A_\mu =\zeta _\mu \phi_{1}(\rho)
e^{ik\cdot x} e^{i\ell \theta}, \quad  k\cdot \zeta = 0, \quad
A_\rho = 0,
  \quad A_\theta = 0\labell{tone}\\
\textrm{Type 2:}&& \quad  A_\mu =0,  \quad  A_\rho = \phi_{2}(\rho)
e^{ik\cdot x} e^{i\ell \theta}, \quad A_\theta = \tilde{\phi}_{2}
(\rho) e^{ik\cdot x} e^{i\ell \theta} \labell{ttwo}\eeqa
where $\ell=0,\pm 1, \pm 2,...$.

For type 1 modes, eq.~\reef{DpGaugeEOM} with $b=\mu$ gives
\beq \label{gaugeT1p}
\partial_ \vr^2 \phi_{1}(\vr) +\frac{1}{\vr} \partial_\vr
\phi_{1}(\vr) + \left( \frac{\bm^2}{(1+\vr^2)^{\frac{7-p}{2}}}
 - \frac{\ell ^2}{\vr^2} \right) \phi_{1}(\vr) =0 \, ,
\eeq
where we have used the redefinitions \reef{mbar}. This equation is
identical to that for the transverse scalars \reef{transp} and
thus the type 1 gauge fields will have the same spectrum as those
scalar modes: $\bm_1(n,|\ell|)=\bm_\perp(n,|\ell|)$, $\ell \ne 0$.

For modes of type 2, eq.~\reef{DpGaugeEOM} with $b=\mu$ gives
\beq
\rho \, \partial_\rho (\rho \phi_{2})= -i\ell \,
\partial_\rho \tilde{\phi}_2 \, . \labell{relphis}
\eeq
There is no regular solution for $\ell=0$ and we only consider $\ell
\ne 0$ in the following analysis. In this case, using
\reef{relphis}, the equations resulting from \reef{DpGaugeEOM} with
$b=\rho$ and $b=\theta$ are equivalent and, using eq.~\reef{mbar},
give
\beq \label{gaugeT2p}
\partial_\vr ^2 \phi_{2} (\vr) + \frac{3}{\vr} \partial_\vr
\phi_{2} (\vr) + \left( \frac{\bm^2}{(1+\vr^2)^{\frac{7-p}{2}}}
 - \frac{\ell ^2-1}{\vr^2}  \right) \phi_{2} (\vr) =0 \, .
\eeq
As usual, for $2\leq p \leq 4$, the spectrum of these fluctuations
follows from solving for $\bm=\bm_2(n,\ell)$ and using \reef{mass}.
These can be matched with the spectrum of `orthogonal' fluctuations
with $\bm_2(n,|\ell|) = \bm_\perp(n,|\ell|)$ for $\ell \ne 0$. The
latter follows since upon substituting $\phi_2=\psi/\vr$,
eq.~\reef{gaugeT2p} reduces to eq.~\reef{transp}.

Thus, the spectrum of mesons in $(p+1)$-dimensional ($2 \leq p\leq
4$) SYM coupled to a fundamental hypermultiplet confined to a
codimension-two defect is discrete with mass gap $M \sim m_q /\gef$.
There are $(7-p)$ scalar mesons corresponding to the `orthogonal'
fluctuations of the probe, two corresponding to `parallel'
fluctuations, one vector corresponding to type 1 gauge fields, and
one scalar corresponding to type 2 gauge fields. The spectra of all
of these are related via:
\beqa M_\perp (n,|\ell|) &=& M_1 (n,|\ell|) = M_2 (n,|\ell|) =
M_-(n,|\ell|+1)=M_+(n,\ell-1)\nonumber\\
&&\quad=M_{\ast ,+}(n,\ell+1)=M_{\ast,-}(n,|\ell|-1) \, .
\labell{DpDpmodes} \eeqa
As usual, the mesons organize themselves into massive
supermultiplets and the masses of the fermionic mesons match those
of the bosonic modes determined here.

Again we have found that the meson masses scale as $M \propto m_q /
\gef(m_q)$, which is {\it identical} to the scaling found in the
previous two cases. Examining the meson spectra of the various brane
configurations in more detail, we note that they are related as the
radial equations for each different probe can be mapped into one
another. Eq.~\reef{transp} for the D$p$-D$p$ brane system reduces to
eq.~\reef{eom2} for the D$p$-D$(p+4)$ system upon substituting:
$\phi_\perp(\vr;n,\ell)=\vr\,\phi(\vr;n,\ell_3=|\ell|-1)$. Hence the
mass levels for these two systems are identical, although the
degeneracies will differ in the two cases. Similarly,
eq.~\reef{transp2} for the D$p$-D$(p+2)$ system maps to
eq.~\reef{eom2} for the D$p$-D$(p+4)$ system upon substituting:
$\phi_\perp(\vr;n,\ell_2)=\sqrt{\vr}\,
\phi\left(\vr;n,\ell_3=\ell_2-{1\over2}\right)$. Of course, this is
only a formal identification because the spectra in each case are
only evaluated for integer values of the angular quantum numbers
$\ell_{2,3}$. Hence the physical spectra differ for these two
systems.

\section{Beyond 10D supergravity: an example} \label{beyond}

It is well understood that, with the exception of $p=3$,  the
supergravity regime discussed above is only applicable for a
well-defined intermediate regime of energy scales  \cite{itzhaki}.
From the results of section \ref{sugraM}, the meson spectra clearly
focus on a very precise energy scale in the field theory, namely,
the quark mass.  Thus, to study meson spectra in gauge theories in
the infrared (IR) or ultraviolet (UV) regimes with this holographic
framework, one must venture beyond ten-dimensional supergravity.

In this section we consider on the meson spectra beyond this
supergravity regime for the specific example of the D2-brane
background, in which we introduce D4- and D2-brane probes. The
corresponding field theory is three-dimensional $U(N)$
super-Yang-Mills (SYM) coupled to a fundamental hypermultiplet on a
codimension-one and -two defects (for the D4- and D2-probes,
respectively) with eight supercharges. For this theory, the
Yang-Mills coupling constant and the dimensionless effective
coupling constant are given by \cite{itzhaki}
\beq \gym^2 = \frac{\gs}{\sqrt{\al'}}\ , \qquad g_{eff}^2(m_q) =
\frac{\gym^2N}{m_q}\ . \eeq

The meson spectrum for the D2-brane theory has different
descriptions depending on the energy  scale $m_q$ \cite{itzhaki}, as
shown in figure \ref{energyFig}. Perturbative SYM is valid for
$g_{eff} \ll 1$ which corresponds to the UV regime or very large
values of $m_q$. The type IIA supergravity description takes over in
the regime $1<g_{eff}<N^{2/5}$, where both the curvature and string
coupling of the gravity background are small. Once $g_{eff} >
N^{2/5}$, the string coupling becomes large and an
eleven-dimensional supergravity description is required. Thus, in
the far infrared, the gravity theory is strongly coupled type IIA
supergravity which lifts to M-theory as an M2-brane background.
\FIGURE{\includegraphics[width=\textwidth]{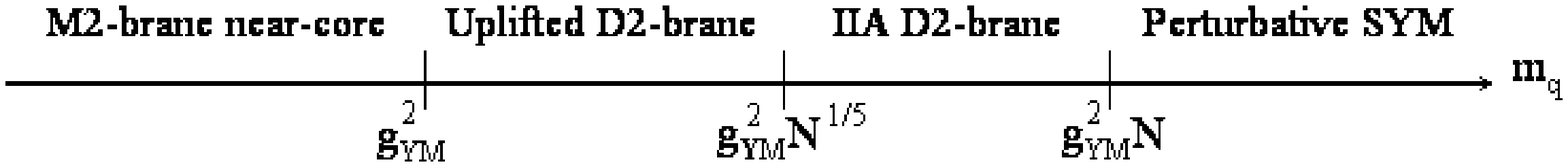}
\caption{Descriptions of the D2-brane theory for different  energy
regimes in the field theory, set by the quark mass $m_q$.}
\label{energyFig}}

The choice of background here was motivated by the desire to compare
our results to those of ref.~\cite{ek}. There an extension of the
gauge/gravity duality beyond the probe approximation with a large
number of fundamental fields  was discussed for a system of D2- and
D6-branes. Their results seem to indicate that the meson mass is
directly proportional to the quark mass, $M\propto m_q$. In
contrast, our results in section \ref{xam} gave $M\propto
m_q^{3/2}$. These disparate scalings are reconciled by noting that
the results of \cite{ek} actually only apply in the far infrared
regime, where the dual description is in terms of the M2-brane
throat rather than the D2 background. We study the theory with D4-
and D2-brane probes instead of D6-branes, because it is easy to
follow the physics of the former probes from the ten-dimensional
regime to the eleven-dimensional phase where the latter lift to M5-
and M2-brane probes, respectively. Our results in the previous
sections indicate that the mass gap in meson spectra of these
different theories should all scale in the same way.

\subsection{Strong coupling: M2-branes with M5- and M2-probes}
\label{strongCoupling}

With $p=2$, eq.~\reef{fullGeom} shows that the dilaton diverges as
$r\rightarrow 0$ and so in the far infrared the system becomes
strongly coupled. As discussed above, the type IIA theory is lifted
to M-theory in this regime. In this case, the array of D2- and
D4-branes in ~\reef{Dp-D(p+2)} or of D2-branes in \reef{Dp-Dp} would
be lifted to a system of $N$ coincident M2-branes with an M5- or
M2-brane probe, as indicated below:
\begin{equation}
\begin{array}{rccccccccccc}
& 0 & 1 & 2 & 3 & 4 & 5& 6&7&8&9&\ 11\\
M2\ & \x & \x & \x &    &    &  &   &  &  &  & \\
M5\ & \x & \x &   & \x & \x  &\x &   &  &  &  &\ \x \\
(probe)\ M2\ & \x &  &   & \x & \x  &  &   &  &  &  & \\
\end{array}\labell{M2-M5}
\end{equation}

The type IIA background solution ~\reef{fullGeom} is readily lifted
to eleven dimensions as (see, for example, \cite{johnson} and
references therein)
\beqa
ds^2 &=& f^{-2/3}(-dt^2 +dx_1^2 +dx_2^2) + f^{1/3}
(dr^2 + r^2 d\Omega_6^2 +d\xii ^2) \labell{M2metric} \\
A_{012}&=&f^{-1}
\eeqa
where $r$ is a radial coordinate in the 3456789-space and $\xii$ is
the compact coordinate in the eleventh dimension with $\xii \sim
\xii +2 \pi \Rii$. Also $f=f(r,\xii)$ is, in general, a harmonic
function of all of the transverse coordinates. Lifting the type IIA
solution \reef{nearGeom} with $p=2$ does not yield exactly the
M2-brane solution but rather the solution for a set of M2-branes
smeared over the circle direction. However, we need to consider the
solution for $N$ coincident M2-branes localized at a point on the
$\xii$-circle. (Of course, even more complicated solutions can also
be constructed.) In this case, the harmonic function is given by
\cite{bigRev}
\beq f(r,x_{11}) = \sum_{n=-\infty}^\infty \frac{2^5\pi^2 \lp^6
N}{[r^2+(\xii  +2\pi nR_{11})^2]^3}\ . \labell{M2f} \eeq
The summation can be carried out to yield a closed-form result
--- see appendix \ref{full} --- but this expression is not very
illuminating for the following.

For the fluctuation analysis, we focus on the limits $r \gg \Rii$
and $r,\xii \ll \Rii$, in which the harmonic function (\ref{M2f})
simplifies considerably. In the large $r$ limit, $r\gg \Rii$, the
dependence on $\xii$ drops out leaving:
\beq
f(r) = \frac{6\pi^2 \lp^6 N}{\Rii}
\frac{1}{r^5}. \labell{upliftD2}
\eeq
In the following, we refer to this as the ``uplifted D2-brane''
solution because this corresponds to precisely the
eleven-dimensional lift of eq.~(\ref{nearGeom}) with $p=2$. Near the
core of the M2-branes ($r,\xii \ll \Rii$), the harmonic function
reduces to
\beq
f(r,\xii) = \frac{2^5 \pi^2 \lp^6 N}{(r^2+\xii ^2)^3}. \labell{nearCore}
\eeq
The metric \reef{M2metric} with this harmonic function gives the
near-horizon geometry for $N$ coincident M2-branes, \ie $AdS_4\times
S^7$ with $N$ units of flux. We will refer to this as the ``near
core'' solution.

In the configuration \reef{M2-M5}, we embed the probe branes at a
distance $L$ from the background M2-branes in, say, the
$X^9$-direction. As the system is supersymmetric, there is no
potential for either of the probe branes and they are in static
equilibrium. Working in the static gauge, we consider small
fluctuations in the 6789-directions transverse to the background
M2-branes
\beq X^A = \delta^A _9\, L +  2\pi \al'\,\chi^A \, , \quad A=6,7,8,9
\eeq
which are overall transverse directions for either probe brane. For
the D2-brane probe, the $\xii$-fluctuations would also fall into
this class.\footnote{Our calculations could be extended to
fluctuations of the M5-brane in the $X^2$ direction (parallel to the
backgroundM2-branes) and of the two-form potential on the M5
worldvolume. Similarly, the M2-brane fluctuations along the $X^1$-
and $X^2$-directions could also be considered. In either case, these
fluctuations couple each other but not to the scalar fluctuations in
the 6789-directions.} For these scalar fields, the relevant
probe-brane action is just the DBI-action for M5-branes \cite{Pasti}
\beq S_{M5} = -\tau_{M5}  \int d^6\xi \sqrt{-\det P[G]_{ab}}\, ,
\eeq
or for M2-branes (see, \eg \cite{DtoM})
\beq S_{M2} = -\tau_{M2}  \int d^3\xi \sqrt{-\det P[G]_{ab}}\, .\eeq
In both cases, $P[G]_{ab}$ denotes the pull-back of the
11-dimensional spacetime metric (\ref{M2metric}) to the probe-brane
worldvolume. The brane tensions are: $\tau_{M5}=(2\pi)^{-5}\lp^{-6}$
and $\tau_{M2}=(2\pi)^{-2}\lp^{-3}$ (see, \eg \cite{johnson}).
Proceeding as in the ten-dimensional analysis above, we expand these
actions to quadratic order in the fluctuations to obtain the
Lagrangian density
\beq \mathcal{L} \simeq -\sqrt{-\det g_{ab}}\left[1+
 \frac{(2\pi\al')^2}{2}f^{1/3} g^{cd}\partial_c\chi^A\partial_d \chi^A \right]
 \ , \labell{lagM2M5}
\eeq
where a sum over $A$ is implicit and $g_{ab}$ is the induced metric
on the probe-brane:
\beqa M5:&&ds^2(g) = f^{-2/3}(-dt^2 + dx_1^2) + f^{1/3}(d\rho^2
+\rho^2 d\Omega_2^2 +d\xii ^2)\, ,\labell{induceM5}\\
M2:&&ds^2(g) = -f^{-2/3}dt^2 + f^{1/3}(d\rho^2 +\rho^2 d\theta^2) \,
. \labell{induceM2} \eeqa

The induced metric factorizes such that its determinant is
independent of the fluctuations.  Furthermore, retaining terms only
to quadratic order in the fluctuations, any dependence on the
fluctuations can be dropped from the factor $f^{1/3}g^{cd}$ in
\reef{lagM2M5}. Hence once again, the quadratic Lagrangian reduces
to a simple free scalar theory in a curved background. The equation
of motion for any one of the fluctuations $\chi^A$ ($A=6,7,8,9$)
then follows from \reef{lagM2M5} as
\beq
\partial_c \left[\sqrt{-\det g_{ab}} f^{1/3}g^{cd}\partial_d \chi  \right] =0 \, .
\labell{eomM5perp} \eeq

Now in accord with the above approximations, we take $r^2=\rho^2
+L^2$ and $f(r, \xii) = f(\rho, \xii)$. It is also convenient to
scale the $\rho,\xii$ coordinates by $L$ to define dimensionless
coordinates $\varrho, \, z$:
\[
\rho = L \vr\ , \quad \xii = Lz\ .
\]
Then, apart from an overall factor, the background parameters only
appear in the harmonic function (\ref{M2fsum}) through the ratio
$L/\Rii$ -- see appendix \ref{full}.

The uplifted D2-brane solution ($r\gg \Rii$) is applicable when $L
\gg \Rii$ where we have
\beq f(\vr) = \frac{6\pi^2 \lp^6 N}{\Rii L^5}
\frac{1}{(1+\vr^2)^{5/2}}\ .\labell{upLiftSimp}  \eeq
In the field theory, this solution corresponds to the regime $\gym
^2 \ll m_q \ll \gym^2 N^{1/5}$ \cite{bigRev}. On the other hand, the
near-core M2-brane solution is relevant for $L\ll \Rii$ where
\beq f(\vr,z) = \frac{2^5 \pi^2 \lp^6 N}{L^6}\frac{1}{(1+\vr^2+z ^2)^3}
\ .\labell{nrCoreSimp}
\eeq
This solution corresponds to even smaller values of the quark mass
in the field theory, \ie $m_q \ll\gym ^2$. In the absence of the
probe branes, the field theory on the worldvolume of the M2-branes
is superconformal.  The probe breaks the conformal invariance
introducing an energy scale $m_q$.

We now compute the spectra of mesons corresponding to the transverse
scalars for each of the probe branes of the uplifted and near-core
geometries. Let us begin with the M5-brane probes. For the uplifted
D2-brane solution (\ref{upLiftSimp}), the harmonic function depends
only on $\vr$, and we proceed via separation of variables:
\beq \chi = e^{ik\cdot x}\, \mathcal{Y}^{\ell_2}(S^2)\,
e^{imLz/\Rii} \phi_{up} (\vr)\ , \quad m=0,\pm 1, \pm 2,...
\labell{ansatzm5up}\eeq
With $M^2=-k^2$ and setting
\beq \quad \bm ^2 = \frac{6\pi^2 \lp^6 N}{\Rii L^3}\, M^2\
,\labell{M2Mbar}  \eeq
the equation of motion \reef{eomM5perp} reduces to the following
radial equation for $\phi_{up}=\phi_{up}(\vr)$
\beq
\partial_\varrho^2\phi_{up} (\vr) + \frac{2}{\vr}\partial_\vr \phi_{up} (\vr) +
 \left[\frac{\bm^2}{(1+\vr^2)^{5/2}}-\frac{\ell_2(\ell_2+1)}{\vr^2}
-\frac{L^2}{\Rii^2}m^2 \right]\phi_{up} (\vr)=0 \, .
\labell{upD2perpEOM} \eeq
Then the meson mass spectrum is:
\beq M^2  = \frac{\bm^2}{6\pi^2}\frac{L^3\Rii}{N\lp^6}=
\frac{4\pi}{3}\bm^2\frac{m_q^3}{\gym^2 N} \, , \labell{M2massUplift}
\eeq
where the dimensionless constants $\bm^2$ are the eigenvalues of
(\ref{upD2perpEOM}). In the above, we have used eq.~\reef{qmass}
and the standard formulae (see, {\it e.g.}, \cite{johnson},
\cite{itzhaki}):
\beq
\Rii = \gs \al'^{1/2}, \quad \lp = \gs^{1/3} \al'^{1/2}, \quad \gym^2 = \gs
\al'^{-1/2}. \labell{defR11}
\eeq

For $m=0$, \ie with no M-theoretic excitations,
eq.~\reef{upD2perpEOM} matches precisely the expected equation
\reef{transp2} for the D2-D4 system and we recover precisely the
same spectrum. In the M-theory context, one can also excite modes
along the $\xii$-circle.  Setting $\Phi
=\phi_{up}(\vr)\mathcal{Y}^{\ell_2}(S^2)$, eq.~\reef{upD2perpEOM}
can be written as
\beq -\nabla^2 \Phi - \frac{\bm^2}{(1+\vr^2)^{5/2}} \Phi =
-\frac{L^2}{\Rii^2} m^2 \Phi \labell{shoe}\eeq
where $\nabla^2$ is the usual Laplacian in three-dimensional
spherical polar coordinates.  Hence eq.~\reef{shoe} has the form of
a three-dimensional Schroedinger equation with potential
$V=-\bm^2/(1+\vr^2)^{5/2}$ and energy eigenvalue $E=-({L}/{\Rii})^2
m^2$. For small $\vr$, the potential approaches $-\bm^2$ while for
large $\vr$, it approaches zero as $-\bm^2 \vr^{-5}$.  Thus, the
lowest energy eigenvalues will be (very) roughly equal to the
minimum of the potential, $-\bm^2$. Hence we would have
$\bar{M}^2\sim m^2\,L^2/\Rii^2$ and the spectrum of such excitations
would be
\beq \widetilde{M}^2  \simeq m^2\frac{L^5}{N\lp^6\Rii}=
m^2\frac{m_q^5}{\gym^6 N}\ . \labell{Mtmass} \eeq
Comparing to eq.~\reef{M2massUplift}, we note that
$\widetilde{M}^2/M^2\sim L^2/\Rii^2$. Recall that the uplifted D2
geometry applies in the regime $L\gg\Rii$ and so here these $m\ne 0$
excitations are extremely heavy and very difficult to excite. As we
move towards $L\sim\Rii$, the harmonic function \reef{M2f} has a
more pronounced dip in the $x_{11}$ direction --- see appendix
\ref{full} --- and the M-theoretic degrees of freedom become
lighter. In the gauge theory, this becomes most pronounced at
$m_q\sim\gym^2$ (\ie $L\sim\Rii$) where instanton effects are
unsuppressed \cite{itzhaki}. The latter reflect the localization of
the M2-brane on the $x_{11}$-circle and so the background goes over
to the near core solution.

For the near core geometry ($L\ll\Rii$), the harmonic function for
our embedding is given in (\ref{nrCoreSimp}). In this case, it is
useful to use spherical coordinates in the (3,4,5,11)-space,
defining the radial coordinate ${\bar{\vr}}$:
\beq {\bar{\vr}}^2 = \vr^2 +z^2\ . \labell{nrCoreRad} \eeq
As usual, we proceed via separation of variables, taking
\beq \chi = e^{ik\cdot x} \ \mathcal{Y}^{\ell_3}(S^3)\,
\phi_{core}({\bar{\vr}}) \, , \labell{nearansatz}\eeq
where $e^{ikx}$ are plane waves in the 01-space and
$\mathcal{Y}^{\ell_3}(S^3)$  ($\ell_3 =0,1,2,...$) are spherical
harmonics on the $S^3$ of unit radius satisfying \reef{S3}. With
$M^2=-k^2$ and now setting
\beq \bm^2= \frac{2^5 \pi^2 N \lp^6}{L^4}\,M^2\ ,
\labell{M2nrCrMbar} \eeq
the full equation (\ref{eomM5perp}) reduces to a radial equation for
$\phi = \phi({\bar{\vr}})$:
\beq
\partial_{\bar{\vr}}^2 \phi_{core}({\bar{\vr}}) +
\frac{3}{{\bar{\vr}}}\partial_{\bar{\vr}}\phi_{core}({\bar{\vr}})
+\left[\frac{\bm^2}{(1+{\bar{\vr}}^2)^3} - \frac{\ell_3(\ell_3
+2)}{{\bar{\vr}}^2}  \right] \phi_{core}({\bar{\vr}}) =0\, . \eeq
The mass eigenvalues can again be determined numerically for this
equation. The mass spectrum of mesons on the codimension-one defect
is now
\beq M^2 = \frac{\bm^2}{2^5 \pi^2}\frac{L^4}{N\lp^6}\
.\labell{MnrCore} \eeq
Clearly, the scaling here has a different form than found in the in
eqs.~\reef{M2massUplift} and \reef{Mtmass} in the uplifted D2
background. Now, with the identification
$m_q=L/(2\pi\al')=L\,R_{11}/(2\pi\ell_p^3)$, this mass scale would
be $M^2\simeq m_q^4/(\gym^4N)$. However, we will argue in the
discussion section that this interpretation is inappropriate  and
that the correct result, appropriate for the superconformal field
theory dual to the AdS$_4\times S^7$ core, is $M^2\simeq m_q^2/N$.

The computation of the spectra in the case of the M2-brane probe is
essentially the same as the M5-brane case. So in the following we
only note the salient differences. In the uplifted limit of the
background, the ansatz \reef{ansatzm5up} is replaced by
\beq \chi = e^{-iM\,t}\,e^{i\ell\,\theta} \tilde{\phi}_{up} (\vr)\ .
\labell{ansatzm2up}\eeq
The fact that the spatial dependence has been reduced to
$\exp(-iM\,t)$ reflects the fact that the fundamental fields are
localized on a codimension-two defect, \ie a point-like defect in
the 2+1 dimensions. Scaling the mass as in eq.~\reef{M2Mbar}, the
radial equation \reef{upD2perpEOM} now becomes
\beq
\partial_\varrho^2\tilde{\phi}_{up} (\vr) + \frac{1}{\vr}\partial_\vr
\tilde{\phi}_{up} (\vr) +
\left[\frac{\bm^2}{(1+\vr^2)^{5/2}}-\frac{\ell^2}{\vr^2}
\right]\tilde{\phi}_{up} (\vr)=0 \, . \labell{upD2perpEOMa} \eeq
Hence the spectrum of bound states scales as in
eq.~\reef{M2massUplift} and in fact, this spectrum exactly matches
that calculated in section \ref{noway} with $p=2$ since
eq.~\reef{upD2perpEOMa} is precisely the same as eq.~\reef{transp}
in this case. Note that since the M2-brane is confined to a point in
the $\xii$-circle, there are no intrinsically M-theoretic degrees of
freedom to be excited at this stage. Rather the transverse scalars
representing the fluctuations in the $\xii$-direction simply match
the type 2 gauge modes in eq.~\reef{tone} (while the type 1 modes of
eq.~\reef{ttwo} do not exist for $p=2$). Of course, this matches the
three-dimensional duality relating the gauge field on the D2-brane
worldvolume to the $\xii$ scalar on M2-brane \cite{DtoM}.

For an M2-brane probing the core limit of the background, the ansatz
\reef{nearansatz} is replaced by
\beq \chi = e^{-iM\,t} \ \mathcal{Y}^{\ell_2}(S^2)\,
\tilde{\phi}_{core}({\bar{\vr}}) \, . \labell{nearansatz2}\eeq
Scaling the mass as in eq.~\reef{M2nrCrMbar}, the new radial
equation is
\beq
\partial_{\bar{\vr}}^2 \tilde{\phi}_{core}({\bar{\vr}}) +
\frac{2}{{\bar{\vr}}}\partial_{\bar{\vr}}\tilde{\phi}_{core}({\bar{\vr}})
+\left[\frac{\bm^2}{(1+{\bar{\vr}}^2)^3} - \frac{\ell_2(\ell_2
+1)}{{\bar{\vr}}^2}  \right] \tilde{\phi}_{core}({\bar{\vr}}) =0
\eeq
and one arrives at mass eigenvalues scaling as in
eq.~\reef{MnrCore}.

\section{Discussion} \label{discus}

In this paper, we used the the gauge/gravity correspondence with
flavour to derive the spectrum of mesons for strongly coupled gauge
theories in various dimensions. In particular, we have studied a
D$k$-brane probe inserted into the near horizon geometry of $N$
coincident D$p$-branes, which in the dual gauge theory corresponds
to having introduced a fundamental hypermultiplet into the
$(p+1)$-dimensional $U(N)$ super-Yang-Mills theory.  The brane
configuration was arranged to be supersymmetric, still preserving
eight of the original sixteen supercharges of the background. For
each system of D$p$- and D$k$-branes considered, we found that the
mesons were deeply bound and the spectra were discrete. Up to
numerical coefficients, the mass gap for all of these theories had a
universal form:
\beq M \sim \frac{m_q}{g_{eff}(m_q)}\ , \quad{\rm where}\ \
g^2_{eff}(m_q) = g_{YM}^2N\,m_q^{p-3} \, . \label{sugraMfinal} \eeq
Here $g_{eff}(m_q)$ is the dimensionless coupling constant in the
gauge theory evaluated at the quark mass $m_q$, which sets the
relevant energy scale for the mesons. The full spectrum is given as
the above mass \reef{sugraMfinal} times dimensionless constants,
which are computed as eigenvalues of an ordinary differential
equation.

These results apply in an intermediate energy regime of the gauge
theory where the dual ten-dimensional supergravity description is
valid. In this regime where $g_{eff} \gg 1$, eq.~\reef{sugraMfinal}
then dictates that the mesons are deeply bound, \ie the meson mass
is much less than (twice) the quark mass. Of course, this result has
a simple interpretation from the bulk point of view. There, a meson
is a bound-state of two strings of opposite orientation
(corresponding to the quark-antiquark pair) stretching between the
probe D$k$-brane and the singularity or horizon at $r=0$. To form a
meson, the ends at the horizon join together to form an open string
on the D$k$-brane probe. The resulting string is much shorter than
the original ones corresponding to the `free quarks', resulting in a
lower energy configuration or a meson with a mass much less than the
quark mass. Since implicitly this discussion depends only on the
background geometry and the positioning of the probe brane, it is
perhaps less surprising that the scaling of the meson masses was
same for all of the different D$k$-branes, even though these
configurations represent very different gauge theories.

In fact, we now argue that the universal form \reef{sugraMfinal} for
the meson masses has a deeper connection to holography. In
investigating general D$p$-brane backgrounds \reef{nearGeom}, one
finds that probing with wave-packets constructed of massless
supergravity fields leads to a `holographic distance-energy'
relation \cite{polpee} which may be expressed as:
\beq E = \frac{U}{g_{eff}(U)}\ , \quad{\rm where}\ \  g^2_{eff}(U) =
g_{YM}^2N\,U^{p-3} \, . \label{holofinal} \eeq
Hence the energy of a such wave-packet propagating in the vicinity
of (the minimal radius of) the D$k$-brane has precisely the same energy
as the excitations of the massless open string fields on the probe
brane. While this may seem a remarkable coincidence, the matching of
the open and closed string energies is crucial to the consistency of
the gauge/gravity duality. Before introducing any fundamental fields
or probe branes, the duality establishes a connection
\reef{holofinal} between the energy scale in the field theory and
the position in the holographic dimension. As emphasized in
\cite{garyjoe}, consistency then requires that physics in the two
dual theories must be local in both of these parameters. In the
presence of probe branes, locality of bulk gravity theory in the
radial direction  is, of course, obviously maintained. Locality of
gauge theory physics in the energy scale was less obvious to begin
with, but our conclusions for the meson masses \reef{sugraMfinal}
establish this nontrivial feature is preserved after the
introduction of fundamental matter. Hence as demanded by holography,
we again find the desired locality on both sides of the duality.

We are lead then to conjecture that in fact holography dictates that the
energy scale of the spectrum of mesons or massless brane fields must
be as given in eq.~\reef{sugraMfinal}. This conjecture may then be
tested in a variety of new contexts. For example, one might consider
the nonsupersymmetric construction of \cite{holoQCD}. There one
finds that if the quark mass is much larger than the QCD scale (\ie
$U_{KK}$) that the meson spectrum takes precisely the desired form.
Of course, for large $m_q$, the system in \cite{holoQCD} is
essentially the D$p$-D$(p+4)$ configuration of section
\ref{mesonsDpDp4} with $p=4$. A more interesting case to consider is
the holographic construction of \cite{sakai} which involved a
nonsupersymmetric configuration of D4-, D8- and $\overline{\rm
D8}$-branes. In this case, the probe brane configuration is very
different from those considered here since the D8-$\overline{\rm
D8}$ system connect in a `wormhole'.\footnote{Ref.~\cite{njl}
recently considered such probe configurations in a supersymmetric
D4-brane background. Ref.~\cite{holoQCD} also discusses a similar
wormhole configuration for a D6-$\overline{\rm D6}$ pair in a D4
background.} However, if one tunes the model, such that the minimal
radius of the wormhole is above the QCD scale, one again finds that
the massive mesons have masses scaling as in eq.~\reef{sugraMfinal}
\cite{sst2,extra}.\footnote{Apart from the massive spectrum, there
is also a set of massless pions whose existence is required by
symmetry-breaking considerations.} This nonsupersymmetric example,
however, highlights one small subtlety. The quarks in the dual
theory are actually massless, by which we mean the `bare' mass in
the UV theory vanishes. However, the infrared dynamics generates a
finite `constituent' quark mass, which can be identified with the
minimal radius of the D8-branes. When we say that the meson spectrum
scales as in eq.~\reef{sugraMfinal} implicitly we are using this
infrared or constituent quark mass. For a general supersymmetric
configuration, the embedding of the probe brane is flat and so there
will be no distinction between the bare and constituent quark
masses. However, we can expect that in nonsupersymmetric
constructions that the bare and constituent quark masses will differ
as a result of bending of the probe --- although perhaps not as
dramatically as in this example. In such cases, it will be the
constituent mass that is the relevant energy scale determining the
holographic scaling of the meson spectrum. As a final comment, we
note that the same energy scale appears in phase transitions for
these systems \cite{temper}.

As indicated above, the full meson spectrum is determined by solving
an ordinary differential equation and imposing regularity
requirements. The masses are then given by the above mass
\reef{sugraMfinal} times dimensionless constants. While in general
one must resort to numerics to determine the proportionality
constants, it was possible to solve the ODE's (in terms of
hypergeometric functions) and hence exact values for the
proportionality constant for $p=3$ --- see appendix \ref{solns}. One
of the outstanding features of these $p=3$ spectra (\ie for D3-D3,
D3-D5, D3-D7) is that they are highly degenerate: all states with
the same $\nu =n +\ell$ have the same mass.  The D3-D7 case was
studied extensively in \cite{spectra} where it was argued that this
degeneracy may arise because of an extra, hidden $SO(5)$ symmetry. A
detailed examination of the spectra  in section \ref{sugraM} shows
that no such degeneracy occurs for $p\ne 3$ --- see, \eg Table
\ref{d2d4}.  Hence it seems that the degeneracy found for $p=3$ is
connected to the conformal invariance of the N=4 super-Yang-Mills
gauge theory.
\TABLE{
\begin{tabular}{c c}
{\begin{tabular}{|c|c|c|c|c|c|}
\cline{3-6}
\multicolumn{2}{c}{ } & \multicolumn{4}{|c|}{$n$} \\
\cline{3-6} \multicolumn{2}{c|}{ } & 0 & 1 & 2 & 3  \\ \hline
     &0    & 4.33 & 23.63 & 56.78 & 103.71 \\
\cline{2-6}
$\ell_2 $ & 1    & 21.02 & 52.11 & 96.90 & 155.42\\
\cline{2-6}
      &  2  & 48.45 & 91.32 & 147.83 & 218.01 \\
\cline{2-6}           &  3      & 86.63 & 141.29 & 209.53 & 291.47 \\
\hline
\end{tabular} }
&
{\begin{tabular}{|c|c|c|c|c|c|c|c|}
\cline{3-6}
\multicolumn{2}{c}{ } & \multicolumn{4}{|c|}{$n$} \\
\cline{3-6}
\multicolumn{2}{c|}{ } & 0 & 1 & 2 & 3  \\
\hline
           &0    & 1.67 & 6.74 & 14.64 & 25.41  \\ \cline{2-6}
$\ell _2$ & 1    & 8.97 & 18.30 & 30.51 & 45.58 \\ \cline{2-6}
           &  2  & 21.47 & 35.06 & 51.53 & 70.89 \\\cline{2-6}
           &  3      & 39.18 & 57.01 & 77.75 & 101.38 \\ \hline
\end{tabular}} \\
$(a)$ & $(b)$
\end{tabular}
\caption{The spectrum for scalar fluctuations in the $(a)$ D2-D4 and $(b)$ D4-D6 brane
configurations in terms of $\bar{M}_\perp ^2(n,\ell_2)$ . }\label{d2d4} }

The connection between the breaking of conformal invariance and of
the degeneracy of the meson spectrum can be made more precise as
follows. If one considers states with fixed $\nu =n +\ell$, one
finds in general that the mass increases (decreases) with increasing
$\ell$ for backgrounds with $p=4$ ($p\le3$) --- as can be seen, \eg
in Tables \ref{d2d4}$a,b$. This trend seems to be connected to the
running of the effective coupling \reef{couple} in the nonconformal
backgrounds. Given that the spectrum scales with $M \sim
m_q/g_{eff}$, the individual masses will depend on the precise value
of the effective coupling which is relevant. For a given state, the
radial profile describes the `structure' of the meson in energy
space and one can calculate the mean energy $\langle U\rangle$ ---
see the technical details in appendix \ref{except}. Tables
\ref{rd2d4}$a,b$ show the results for the D2-D4 and D4-D6 systems,
which are of the general case. Note that $\langle U \rangle$
increases when the radial quantum number $n$ is increased but
decreases with increasing $\ell_2$.\footnote{The latter is an
interesting effect. The `angular momentum' produces a repulsive,
centrifugal barrier, causing the peak of the wave function to move
away from the origin. However, as $\ell_2$ is increased, the
asymptotic potential is also raised causing the asymptotic
wavefunction to fall off more quickly. The latter is a stronger
effect so that the wavefunction localizes at smaller radii and hence
smaller $\langle U\rangle$, as seen in tables \ref{rd2d4}$a,b$.}
Hence combining the facts that the effective coupling grows with
increasing energy (radius) in the D4 background and that the meson
masses scale (roughly) inversely with $g_{eff}$, the meson masses
would be expected to increase more with an increase in $\ell$ than
with a corresponding increase in $n$. Of course, the reverse trend
should be expected for the D2 background, where the effective
coupling decreases at larger energies. Hence this reasoning
reproduces the trend in the mass spectra commented on above and
illustrated in Table \ref{d2d4}. This phenomenon is similar to the
effects seen in the spectroscopy of heavy-quark mesons due to the
running of the QCD coupling \cite{real}.
\TABLE{
\begin{tabular}{c c}
{\begin{tabular}{|c|c|c|c|c|}
\cline{3-5}
\multicolumn{2}{c}{ } & \multicolumn{3}{|c|}{$n$} \\
\cline{3-5}
\multicolumn{2}{c|}{ } & 0 & 1 & 2   \\
\hline
 & 0    & 1.4173 & 1.5764 & 1.6541  \\ \cline{2-5}
     $\ell_2 $  & 1    &  1.3594  & 1.4620 & 1.5278   \\\cline{2-5}
           &  2      & 1.3375 & 1.4130 & 1.4676  \\ \hline
\end{tabular}}
&
{\begin{tabular}{|c|c|c|c|c|}
\cline{3-5}
\multicolumn{2}{c}{ } & \multicolumn{3}{|c|}{$n$} \\
\cline{3-5}
\multicolumn{2}{c|}{ } & 0 & 1 & 2   \\
\hline
 & 0    & 2.9193 & 5.9310 & 8.8394  \\ \cline{2-5}
     $\ell_2 $  & 1    &  2.1634  & 3.1276 & 4.0868   \\\cline{2-5}
           &  2      & 1.9985 & 2.5708 & 3.1427  \\ \hline
\end{tabular}} \\
$(a)$ & $(b)$ \\
\end{tabular}
\caption{The radial expectation value $2\pi \langle U\rangle/m_q$
for the scalar field $\chi$ in the $(a)$ D2-D4 and $(b)$ D4-D6 brane systems.
}\label{rd2d4}}

Ref.~\cite{fifth} calculated the form factors of the mesons in the
D3-D7 system with respect to various conserved currents. They found
that the size of these bound states to be roughly $1/M \sim \sqrt{
\gym^2 N}/m_q$. From the point of view of supergravity, this result
is not difficult to understand: the form factors are determined by
the overlap of various radial profiles and $M$ is the only scale in
the problem. Hence it is the only dimensionful quantity that can set
the size of the mesons. This analysis can be extended to mesons in
the general D$p$-D$k$ systems considered here and the same behaviour
will be found with the meson size set by $1/M \sim \gef(m_q)/ m_q$.
It would be interesting to study these issues in more detail.

As noted in section \ref{sugraM}, we noted that the analysis of the
meson spectrum fails for $p=5$ because there are no normalizable
solutions. This is related to the usual problematic features of
holography for D5-branes \cite{itzhaki}. At large radius, the
coupling becomes large and the appropriate gravity background is
that of the S-dual NS5-brane, where the holographic dual is
$\cal{N}$=(1,1) little string theory --- see, \eg \cite{lst1,lst}.
The latter background includes an infinite throat containing
delta-function normalizable states and so it is natural that
analogous `meson' states would arise when a probe brane is
introduced. This would be an interesting topic for further study.

The computations of the $p=2$ spectra in section \ref{beyond} were
motivated by the comparsion with previous results for holographic
mesons in (2+1)-dimensions \cite{ek}. The latter went beyond the
probe approximation and considered a holographic description of a
large number of flavours in the strongly coupled D2-D6 system. In
particular, their construction began with $N_c$ D2-branes and $N_f$
D6-branes alligned as shown in figure \reef{D2-D6} and used the
fully back-reacted near-horizon geometry of \cite{pelc} in the limit
$1 \ll N_f < N_c$ but with $N_f/N_c$ finite. In this limit with the
back-reacted geometry, the fluctuations of the D6-brane correspond
to fluctuations in the supergravity background and so mesons are
described by closed string states in the bulk. Ref.~\cite{ek}
attempted to solve for meson masses numerically and found a discrete
spectrum with a mass gap $M \propto m_q$. The latter seems to
contradict the results in section \ref{xam} where we found that the
meson masses scale as $M\propto m_q^{3/2}/g_{YM}^2 N$. This
disagreement is simply resolved by realising that the two
computations are actually performed for different energy regimes in
the field theory, as illustrated in figure \ref{energyFig}. Our
ten-dimensional supergravity result holds in the intermediate energy
$\gym^2N^{1/5}\ll m_q\ll\gym^2N$. On the other hand, \cite{ek}
consider the far infrared limit $m_q\ll\gym^2$, where the gravity
description becomes an M-theory configuration, \ie an M2-brane
background with orbifold identifications \cite{pelc}.

To study meson spectra in different holographic regimes, we turned
in section \ref{beyond} to the D2-D4 and D2-D2 systems for which
probe brane computations are possible at both the intermediate and
far infrared energy scales. These correspond to defect theories but
the results of section \ref{sugraM} show that the mass scale of
meson spectra remains the same in these cases. In the infrared
regime, the D2-brane lifts to an M2-brane and the D4- and D2-brane
probes lift to M5- and M2-branes, respectively. In the uplifted
D2-brane regime, the masses of the lowest lying mesons scale as
$M^2\propto L^3R_{11}/(N_c\ell_p^6)\sim m_q^{3/2}/(g_{YM}^2 N)$,
which matches that found in the ten-dimensional regime. However, we
see in eq.~\reef{MnrCore} that this scaling changes in the near-core
regime: $M^2 = L^4/(N_c\lp^6)$. This scaling actually precisely
matches that found in \cite{ek} (once the AdS$_4$ curvature,
$R^6_{AdS}=N_c\lp^6$, is restored). This agreement is almost better
than we might have hoped for given that our probe approximation
applies for $N_f/N_c=0$ whereas \cite{ek} has $N_f/N_c$ small but
finite.

As explained in the previous section, if we tentatively identified
$m_q=L/(2\pi\al')=L\,R_{11}/(2\pi\ell_p^3)$ then the mass scale
\reef{MnrCore} in the near core regime would become $M^2\simeq
m_q^4/(\gym^4N)$ with an extra factor of $m_q/\gym^2$ compared to
the intermediate regime where ten-dimensional supergravity is valid.
However, this is clearly an inappropriate interpretation of the
result \reef{MnrCore} which describes the meson spectra in the far
infrared where the gauge theory has flowed to the superconformal
fixed point. In particular, the latter no longer has any
knowledge of the scales $R_{11}$ or $\gym$. The introduction of a
probe brane in the AdS$_4\times S^7$ background corresponds to
coupling the SCFT to massive fundamental fields, which break the
conformal invariance (mildly) through their mass. This mass scale
must be set by $L$, the only available scale in the supergravity
description. Certainly one should not expect that one can extend
naively extend the identification $m_q\propto L$ found in ten
dimensions to the near core regime. In particular, $L$ is now an
arbitrary coordinate distance in the core region. Instead we must
re-evaluate using the standard distance-energy relation for AdS$_4$
\cite{polpee}, which associates the energy with the `standard' AdS
coordinate $U$. Then with the appropriate coordinate transformation,
the mass of the fundamental fields in this regime becomes $m_q\simeq
L^2/\ell_p^3$, which matches the choice made in \cite{ek}. The mass
scale in the meson spectrum is then $M\simeq m_q/N_c^{1/2}$. Again
this result matches the desired `holographic distance-energy'
relation \cite{polpee}. That is, the scaling of the meson spectrum
here once again conforms to the `energy locality' demanded by
holography, as discussed above.

It is also natural to consider the extension of the $p=4$ theories
to the strong coupling M-theory regime, as well. In this case, the
dilaton grows with radius in the D4 background and one lifts the the
M5-brane background at very high energies. In the dual description,
this corresponds to the UV completion of the original
five-dimensional gauge theory being a six-dimensional
$\cal{N}$=(0,2) SCFT (on a circle) \cite{moshe}. If we introduce
fundamental fields on a codimension two defect as in section
\ref{noway} with $p=4$, the D4 probe also lifts to an M5-brane
(wrapping the eleventh dimension). The fundamental matter
excitations are now naturally strings associated with M2-branes
ending on the probe M5-brane. The low-lying `mesons' again
correspond to fluctuations of the massless fields on the M5 probe
and one finds that the mass gap is again given by $m_q/N^{1/2}$, in
agreement with the `holographic distance-energy' relation
\cite{polpee}.

As commented above, the agreement of our probe calculations in the
AdS$_4$ core of the D2-brane background yielded remarkable agreement
with the results at finite $N_f/N_c$ in \cite{ek}. It would be
interesting to pursue this further in other backgrounds. Following
the results of \cite{local}, there have been several attempts
\cite{NfNc,NfNc2} to establish fully back-reacted backgrounds
describing four-dimensional gauge theories at finite $N_f/N_c$. It
seems further work is needed to develop these solutions to the point
where the meson spectra could be calculated. However, some simple
observations can be made at this stage. In some approaches
\cite{NfNc2}, the back-reacted solution still includes flavour
branes. However, in other models {ek}, the branes are replaced by
deformations of the supergravity background and so the open-string
excitations representing the mesons would be replaced by
closed-string states. On general grounds, the latter are expected to
satisfy the `holographic distance-energy' relation in these new
backgrounds. Hence, from this point of view, it seems that
holography will dictate the same kind of scaling in the meson
spectra at both finite and vanishing $N_f/N_c$. Hence we expect the
`remarkable' agreement observed here for $p=2$ might extend to more
general situations.

One direction for future investigation could be considering mesons
with higher spin in the $(p+1)$-dimensional gauge theories. This
would require a straightforward extension of the calculations for
the D3-D7 system in \cite{spectra}. There  the spectrum for large
$J$ was calculated from large semiclassical strings hanging down
from the probe brane. These spinning string solutions can also be
extended to consider bound states of quarks with different masses by
introducing separated probe branes \cite{multi}. One universal
result will be that in analogy to \cite{spectra}, the spectrum will
start with a Regge-like trajectory where the tension is governed by
the quark mass. In this regime, the gravity description corresponds
to spinning strings whose extent is (larger than the string scale
but) smaller than the curvature scale of the background. The Regge
slope is then just the redshifted tension of fundamental string and
in field theory parameters is given by $1/\al'_{\rm eff}\simeq
m_q^2/\gef$. While easily understood from the supergravity point of
view, this behaviour is surprising from a field theory point of view
as none of the theories considered here are confining \cite{regee}.
For $J \gg \gef$, the size of string exceeds the background
curvature scale. In this regime, it should be that the mass spectrum
can be understood as that of two nonrelativistic and widely
separated quarks which are weakly bound by a long-range potential,
$V\simeq-(\gym^2N/L^2)^{1/5-p}$ --- the latter being that computed
for two static quarks in the corresponding background
\cite{Maldacena98}.

\acknowledgments

We wish to thank Roberto Casero, Steve Godfrey, Jaume Gomis, Jordan
Hovdebo, Harry Lam, Martin Kruczenski, Peter Langfelder, Angel
Paredes, Nemani Suryanarayana and especially David Mateos for
discussions and comments. Research at the Perimeter Institute is
supported in part by funds from NSERC of Canada and MEDT of Ontario.
Research by RCM is further supported by an NSERC Discovery grant and
RMT by an NSERC Canadian Graduate Scholarship. RMT would like to
thank the organizers of {\it Strings05} for the opportunity to give
a poster presentation of this work.

\appendix

\section{Analytic meson spectra for $p=3$}\label{solns}

In section \ref{sugraM} we studied the spectra of mesons
corresponding to fluctuations of probe D$k$-branes in the
near-horizon geometry of $N$ coincident D$p$-branes.  For each brane
configuration and each type of fluctuation of the probe brane, we
found an ordinary differential equation for the radial profile.  As
discussed in section \ref{mesonsDpDp4}, the solutions of these
differential equations must be real-valued, regular and normalizable
in order to be dual to a physical meson state in field theory. Thus,
solutions were chosen that were real and regular at the origin.  The
eigenvalues $\bm$ were then determined by requiring that the
solutions be convergent as $\rho \to \infty$.  The meson spectrum
was then given by \reef{mass}. Though it was not possible to solve
the differential equations analytically for general $p$, analytic
solutions are possible for $p=3$ and we present these solutions and
the resulting meson spectra in this appendix.

The D3-D7 brane system was studied in \cite{spectra} and we review
the analytic solutions and meson spectra briefly here.  The branes
were oriented as shown in the following array:
\begin{equation}
\begin{array}{ccccccccccc}
 & 0 & 1 & 2 & 3 &4 & 5 & 6 & 7 & 8 & 9\\
D3 & \x & \x & \x &  \x &   & & &  &  & \\
D7 & \x & \x & \x & \x & \x  &\x &\x & \x &  &   \\
\end{array}\labell{D3-D7}
\end{equation}
The D7-brane fluctuates in the 89-directions and the radial ODE for
these scalar modes is given in \reef{eom2}. Defining
\beq
\al = - \frac{1}{2} + \frac{1}{2} \sqrt{1+\bm^2} \geq 0,
\eeq
the solution for the radial function $\phi(\rho)$ is
\beq \label{D3D7scalar}
\phi(\rho ) = \rho^{\ell_3} (\rho^2 +L^2) ^{-\al} \,
F(-\al \, , \, -\al + \ell_3 +1\, ;\, \ell_3 +2 \, ; \, -\rho^2/L^2)\, ,
\eeq
where $F(a,b;c;y)$ is a hypergeometric function satisfying (see, \eg
\cite{wyld})
\beq y(1-y) \, u''(y) + [c-(a+b+1)y] \, u'(y) - ab \, u(y) =0. \eeq
To determine the behaviour of solutions as $\rho \to \infty$, the
following asymptotic expansion of $F(a,b;c;y)$ is useful
\cite{wyld}:
\beq
\lim_{y \to \infty} F(a\, ,b\, ;\,c\,;\,y)  =
\frac{\Gamma(c)\Gamma(b-a)}{\Gamma(b)\Gamma(c-a)}(-y)^{-a}
+ \frac{\Gamma(c)\Gamma(a-b)}{\Gamma(a)\Gamma(c-b)}(-y)^{-b}.
\eeq
Using this asymptotic expansion, the regularity of
\reef{D3D7scalar} is achieved by setting $-\al +\ell_3 + 1 =-n$
($n=0,1,2,...$).  The solution is then
\beq \label{D3D7scalar2}
\phi(\rho ) = \frac{\rho^{\ell_3}}{(\rho^2 +L^2) ^{n+\ell_3+1}}
\, F(-(n+\ell_3+1)\, ,\,-n\,; \, \ell_3 +2 \,; \, -\rho^2/L^2) \, ,
\eeq
with mass eigenvalues
\beq
\bm^2_s = 4(n+\ell_3 +1)(n+\ell_3 +2) \, .
\eeq
The solutions for the gauge fields are determined similarly. The ODE
for the type 1, 2, and 3 modes are given in eqs.~\reef{Iphi},
\reef{DpT2}, and \reef{DpT3}, respectively. The solutions are
\beqa \phi_1(\rho) &=& \rho^{\ell_3} (\rho^2 +L^2) ^{-\al}
F(-\al + \ell_3 +1 \, , \, -\al\, ; \, \ell_3 +2\, ; \, -\rho^2/L^2) \, ,
\labell{abcx1}\\
\phi_2(\rho) &=& \rho^{\ell_3-1} (\rho^2 +L^2) ^{-\al}
F(-\al + \ell_3 +1 \, , \, -\al\, ; \, \ell_3 +2\, ; \, -\rho^2/L^2) \, ,
\nonumber\\
\phi_{3,+}(\rho) &=& \rho^{\ell_3+1} (\rho^2 +L^2) ^{-1-\al}
F(-\al + \ell_3 +2 \, , \,-1 -\al\, ; \, \ell_3 +2\, ; \, -\rho^2/L^2) \, ,
\nonumber\\
\phi_{3,-}(\rho) &=& \rho^{\ell_3+1} (\rho^2 +L^2) ^{-1-\al} F(-\al
+ \ell_3 \, , \,1 -\al\, ; \, \ell_3 +2\, ; \, -\rho^2/L^2) \, ,
\nonumber \eeqa
with spectra
\begin{equation}
\begin{array}{lcc}
\bm^2_1 = 4(n+\ell_3+1)(n+\ell_3+2) \, , \quad & n\geq 0 \, ,  & \ell_3 \geq 0 \, ; \\
\bm^2_2 = 4(n+\ell_3+1)(n+\ell_3+2)  \, , \quad& n\geq 0 \, , & \ell_3 \geq 1 \, ; \\
\bm^2_{3,+}  =  4(n+\ell_3+2)(n+\ell_3+3) \, , \quad & n\geq 0 \, , & \ell_3 \geq 1 \, ; \\
\bm^2_{3,-}  =  4(n+\ell_3)(n+\ell_3+1) \, ,  \quad & n\geq 0 \, , & \ell_3 \geq 1 \, . \\
\end{array}
\end{equation}
We can see explicitly here that the spectra of the various modes are
related as in eq.~\reef{DpDp2modes}. As noted in \cite{spectra}, all
states with the same $n+\ell_3$ have the same mass and so the meson
spectrum has a large degeneracy.

The D3-D5 brane system was oriented as follows:
\begin{equation}
\begin{array}{ccccccccccc}
 & 0 & 1 & 2 & 3 &4 & 5 & 6 & 7 & 8 & 9\\
D3 & \x & \x & \x &  \x &   & & &  &  & \\
D5 & \x & \x & \x &  & \x  &\x &\x &  &  &   \\
\end{array}\labell{D3-D5}
\end{equation}
In this system, there are several classes each of scalar and gauge
field fluctuations.  First, the scalar fluctuations of the D5-brane
may be parallel to the background D3-branes along the $X^4$
direction or orthogonal to the D3-branes in the 789-directions. The
gauge field modes are characterized as type 1, 2 and 3, where the
type 3 excitations couple to the parallel scalar modes. The ODE for
the orthogonal scalars and the type 1 gauge fields are identical and
are given in eqs.~\reef{transp2} and \reef{gaugeT1p2}, respectively.
The solutions for these modes are
\beq \phi_\perp (\rho) = \phi_1(\rho) = \rho^{\ell_2} (\rho^2
+L^2)^{-\al} \, F(-\al + \ell_2+1/2 \,, \,-\al \,; \, \ell_2+3/2 \,
;\, -\rho^2/L^2) \, , \labell{abcx2}\eeq
and the resulting spectra are
\beq
\bm^2_{\perp} = \bm^2_{1} = 4(n+\ell_2+1/2 )(n+\ell_2 +3/2)
\, , \qquad n\geq 0 \, , \quad \ell_2 \geq 0\, .
\eeq
The type 2 gauge fields satisfy eq.~\reef{gaugeT2p2} and the
solutions are
\beq
\phi_2 (\rho) =  \rho^{\ell_2-1} (\rho^2 +L^2)^{-\al} \,
F(-\al + \ell_2+1/2 \,, \,-\al \,; \, \ell_2+3/2 \, ;\, -\rho^2/L^2) \, ,
\eeq
with the corresponding spectrum
\beq
\bm^2_2= 4(n+\ell_2+1/2 )(n+\ell_2 +3/2) \, ,
\qquad n\geq 0 \, , \quad \ell_2 \geq 1\, .
\eeq
For $\ell_2 =0$, the parallel fluctuations are not coupled to the
gauge fields and the ODE reduces to eq.~\reef{parSclr}. The solution
is
\beq
\phi_{||} (\rho) =   (\rho^2 +L^2)^{\al} \,
F(\al \,, \,\al +5/2  \,; \, 3/2 \, ;\, -\rho^2/L^2) \, ,
\eeq
with masses
\beq
\bm^2_{||}= 4(n+3/2 )(n+5/2) \, ,
\qquad n\geq 0 .
\eeq
The type 3 gauge fields and the parallel scalar fluctuations are
coupled for $\ell_2 \geq 1$.  The diagonalized differential
equations were given in eq.~\reef{coupledEOM} with solutions
\beqar
\tilde{\phi} _+ (\rho) &=& \rho^{\ell_2+1}(\rho^2+L^2)^{-\al-1}
\, F(-\al+\ell_2+3/2 \, , \, -\al-1 \, ; \, \ell_2+3/2\, ; \, -\rho^2/L^2) \, \\
\tilde{\phi} _- (\rho)& =& \rho^{\ell_2+1}(\rho^2+L^2)^{-\al-1}
F(-\al+\ell_2-1/2 \, ,  \, -\al+1 \, ;  \, \ell_2+3/2 \, ; \, -\rho^2/L^2) \, ,
\eeqar
and spectra
\begin{equation}
\begin{array}{lcc}
\bm^2_+ =  4(n+\ell_2+3/2)(n+\ell_2+5/2) \, , \quad& n\geq 0 \, , & \ell_2 \geq 1 \, ; \\
\bm^2_- = 4(n+\ell_2-1/2) (n+\ell_2+1/2) \, , \quad & n\geq 0 \, ,  & \ell_2 \geq 1 \, .
\end{array}
\end{equation}
The spectra for the various modes satisfy the relations in
eq.~\reef{DpDp2modes} and there is again a large degeneracy as the
masses only depend on $n+\ell_2$.

Finally, the D3-D3 brane system was oriented as shown in the following array:
\begin{equation}
\begin{array}{ccccccccccc}
 & 0 & 1 & 2 & 3 &4 & 5 & 6 & 7 & 8 & 9\\
\textrm{background} & \x & \x & \x &  \x &   & & &  &  & \\
\textrm{probe} & \x & \x &  &  & \x  &\x & &  &  &   \\
\end{array}\labell{D3-D3}
\end{equation}
Again, the scalar and gauge field fluctuations fall into several
classes. The D3-brane probe fluctuates in the $X^2$ and $X^3$
directions parallel to the background branes and also in the
6789-directions orthogonal to the background branes.  The radial
ODE's for the orthogonal fluctuations \reef{transp} and for type 1
gauge fields \reef{gaugeT1p} were identical.  The solutions for
these fields are
\beq \phi_\perp (\rho) = \phi_1(\rho) = \rho^{|\ell|} (\rho^2
+L^2)^{-\al} \,  F(-\al + |\ell| \, , \,  -\al \, ; \,  |\ell|+1 \,
; \, -\rho^2/L^2) \, , \labell{abcx3}\eeq
and with spectra
\beq
\bm^2_{\perp} = \bm^2_{1} = 4(n+|\ell|)(n+|\ell| +1)\,  ,
\quad n\geq 0 \, ,  \quad  |\ell| \geq 1 \, .
\eeq
The diagonalized radial equation for $\Psi$ corresponding to the
coupled $X^{2,3}$-scalars appears in eq.~\reef{scalarParP}.  The $\ell \ge 0$ and
$\ell <-1$ solutions\footnote{Recall from footnote \ref{lminus1} that supersymmetry
requires the $\ell=-1$ mode to vanish.} are, respectively,
\beqar
\phi_{||,+}(\rho) &=& \rho^{\ell} (\rho^2 +L^2)^{-\al-1}
\, F(-\al -1 \, , \,  -\al+\ell+1 \, ;  \, \ell+1 \, ;  \, -\rho^2/L^2) , \\
\phi_{||,-}(\rho) &=& \rho^{|\ell|} (\rho^2 +L^2)^{-\al-1}
\, F(-\al + 1 \, , \,  -\al+|\ell|-1 \, ;  \, |\ell|+1 \, ;  \, -\rho^2/L^2) \,\, .
\eeqar
with the corresponding spectra
\begin{equation}
\begin{array}{lcc}
\bm^2_{+} =  4(n+\ell +1)(n+\ell +2) \, ,
\quad & n\geq 0 \, ,  & \ell \geq 0; \\
\bm^2_{-} =  4(n+|\ell|-1)(n+|\ell|) \, ,
\quad& n\geq 0 \, , & \ell <-1 \, .
\end{array}
\end{equation}
Since the ODE defining the spectrum for the $\Psi^\ast$ modes is
eq.~\reef{scalarParP} with $\ell \to -\ell$, the spectrum for these
modes is:
\begin{equation}
\begin{array}{lcc}
\bm^2_{\ast, -} =  4(n+|\ell|+1)(n+|\ell|+2) \, ,
\quad & n\geq 0 \, ,  & \ell \leq 0  ; \\
\bm^2_{\ast, +} =  4(n+ \ell -1)(n+ \ell ) \, ,
\quad& n\geq 0 \, , & \ell >1 \,.
\end{array}
\end{equation}
For type 2 gauge fields, the radial equation was given in
eq.~\reef{gaugeT2p} and has solutions
\beq
\phi_{2}(\rho)= \rho^{|\ell|-1} (L^2 + \rho^2)^{-\al}
 \, F(-\al +|\ell| \, ,  \, -\al \, ;  \, |\ell|+1 \, ;  \, -\rho^2/L^2) \,  ,
\eeq
and the resulting spectrum is
\beq
\bm^2_{2}= 4(n+|\ell|)(n+|\ell|+1) \,  , \quad n\geq 0 \, , \quad |\ell| \geq 1 \, .
\eeq
The spectra of these bosonic modes are related as in
eq.~\reef{DpDpmodes} and as the masses depend only on $n+|\ell|$,
the spectrum again exhibits a large degeneracy.

\section{Localized M2-brane background} \label{full}

In section \ref{strongCoupling}, we studied an M5-brane probe in the
near-horizon geometry of $N$ coincident M2-branes. Here, we briefly
discuss some details about the near-horizon geometry of the
M2-branes.

The metric for $N$ coincident M2-branes is (see, \eg \cite{johnson}
and references therein)
\beq ds^2 = f^{-2/3}(-dt^2 +dx_1^2 +dx_2^2) + f^{1/3}(dr^2 +r^2
d\Omega_6^2 +d\xii ^2) \labell{M2metric2} \eeq
where $r$ is a radial coordinate in the 3456789-space and $\xii$ is
the coordinate in the eleventh dimension.  We take the eleventh
dimension to be compact, with $\xii = \xii + 2 \pi \Rii$. For
simplicity, we consider the harmonic function for the case when all
$N$ M2-branes are localized at some point along $\xii$ --- of
course, more complicated configurations are possible. The harmonic
function for the M2-branes can simply be written as a sum over
images in the $\xii$-direction \cite{bigRev}:
\beq f(r,x_{11}) = \sum_{n=-\infty}^\infty \frac{2^5\pi^2 \lp^6
N}{[r^2+(\xii +2\pi nR_{11})^2]^3}. \labell{M2f22} \eeq
It is relatively straightforward to sum this expression in closed
form using techniques from complex analysis (see, \eg
\cite{compact}). The final result can be written as:
\beqa
f(r,\xii) &=& \frac{2\pi ^2 \lp^6 N}{\Rii ^6}
\left[3\frac{\Rii^5}{r^5}\frac{\sinh (r/\Rii)}{\cosh(r/\Rii)- \cos
(\xii/\Rii)} + 3\frac{\Rii^4}{r^4}
\frac{\cosh(r/\Rii)\cos(\xii/\Rii) -1}{(\cosh(r/\Rii)-\cos
(\xii/\Rii))^2}
\right. \nonumber \\
&&  \left.- \frac{\Rii^3}{r^3} \frac{\sin
(r/\Rii)(2-\cos^2(\xii/\Rii)
-\cosh(r/\Rii)\cos(\xii/\Rii))}{(\cosh(r/\Rii)-\cos (\xii/\Rii))^3}
\right]. \labell{M2fsum} \eeqa
As in the fluctuation analysis in section \ref{beyond}, we take
$r^2=\rho^2 +L^2$ and scale the $\rho$ and $\xii$ coordinates by $L$:
\[
\rho = L\, \vr\ , \qquad \xii = L\,z\ ,
\]
where the $z$-periodicity is given by $z \sim z +2\pi \Rii/L$. The
harmonic function (\ref{M2fsum}) can then be expressed as
\beqa
f(\rho,z) &=& \frac{2\pi ^2 \lp^6 N}{\Rii ^6}
\left[3\frac{\Rii^5}{L^5} \frac{1}{(1+\vr^2)^{5/2}}\frac{\sinh
(\frac{L}{\Rii} \sqrt{1+\vr^2})}{\cosh(\frac{L}{\Rii}
\sqrt{1+\vr^2}))-\cos (\frac{L}{\Rii}z)}
\right. \nonumber\\
&&+
3\frac{\Rii^4}{L^4}\frac{1}{(1+\vr^2)^{2}}\frac{\cosh(\frac{L}{\Rii}
\sqrt{1+\vr^2}))\cos(\frac{L}{\Rii}z) -1}{(\cosh(\frac{L}{\Rii}
\sqrt{1+\vr^2}))-\cos (\frac{L}{\Rii}z))^2} \labell{M2fsumSc}\\
&&  \left.-\frac{\Rii^3}{L^3}\frac{1}{(1+\vr^2)^{3/2}} \frac{\sin
(\frac{L}{\Rii} \sqrt{1+\vr^2}))(2-\cos^2(\frac{L}{\Rii}z)-
\cosh(\frac{L}{\Rii} \sqrt{1+\vr^2}))\cos(\frac{L}{\Rii}z))}
{(\cosh(\frac{L}{\Rii} \sqrt{1+\vr^2}))-\cos (\frac{L}{\Rii}z))^3}
\right]\nonumber. \eeqa

In section \ref{strongCoupling} we focused on the regimes $L \gg
\Rii$ and $L \ll \Rii$, which we referred to as the ``uplifted
D2-brane'' and ``near core'' solutions, respectively.  Figure
\ref{fPlots} displays plots of the full function
\reef{M2fsumSc} for various values of $L/\Rii$ to illustrate how the
full harmonic function changes in different parameter regimes.
Figure \ref{fPlots}a displays the harmonic function for $L / \Rii
=50$. This illustrates how in the $L\gg \Rii$ regime the structure
in the $\xii$ direction is washed out and $f$ essentially depends
only on the radial coordinate $\rho$. That is, the full harmonic
function is well-approximated by the uplifted D2-brane solution
\reef{upLiftSimp}.  The harmonic function in the limit $L \ll \Rii$
is shown in figure \ref{fPlots}c, where $L/\Rii = 10^{-4}$. In
this case, the structure in the $\rho$ and $z$ coordinates is the
same and $f(\rho, z)$ can be approximated by the near core solution
\reef{nearCore}.  Figure \ref{fPlots}b shows the harmonic function
for $L/\Rii =3$, an intermediate regime.
\FIGURE{
 \begin{tabular}{ccc}
 \includegraphics[width=5cm]{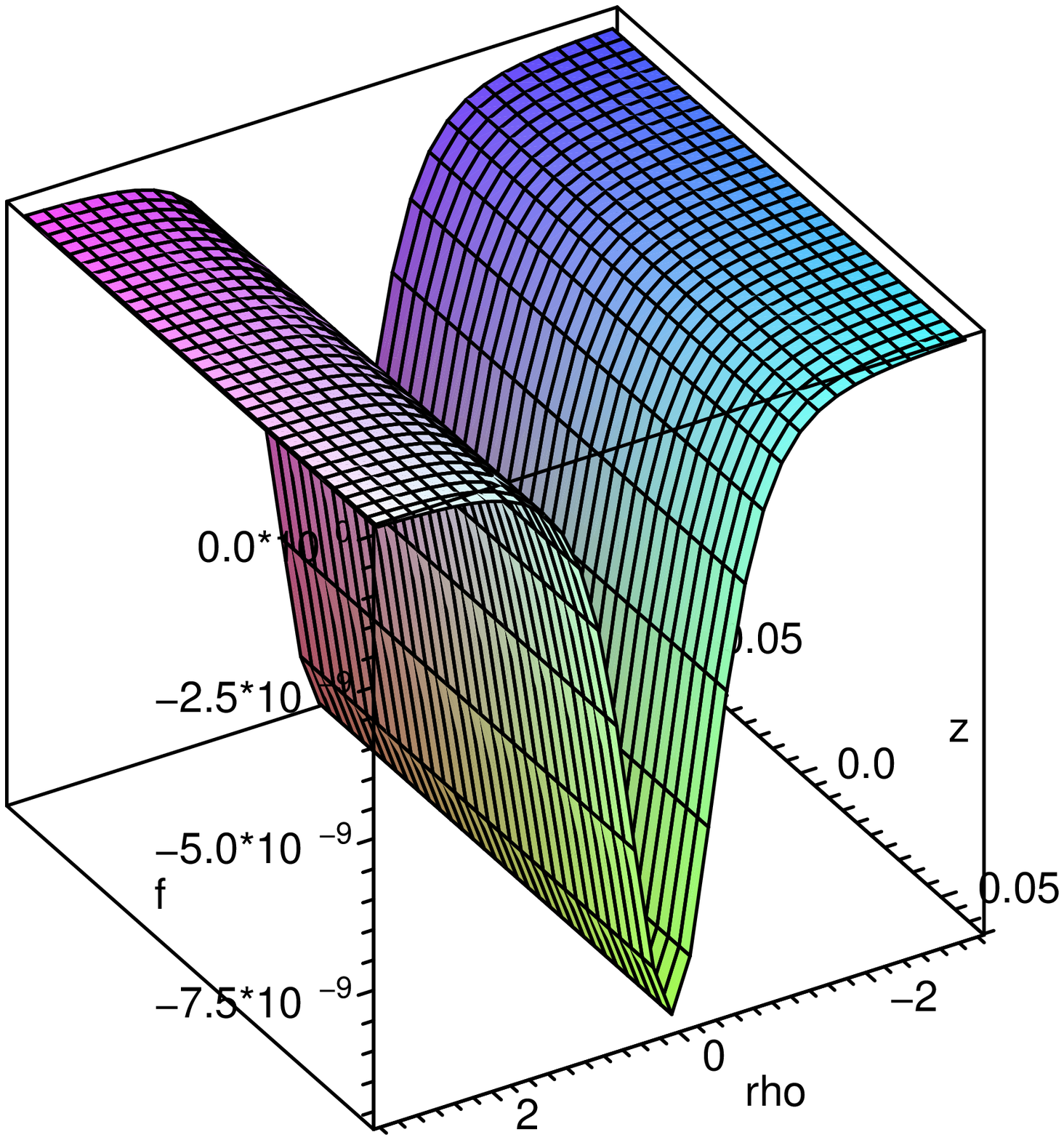} &  \includegraphics[width=5cm]{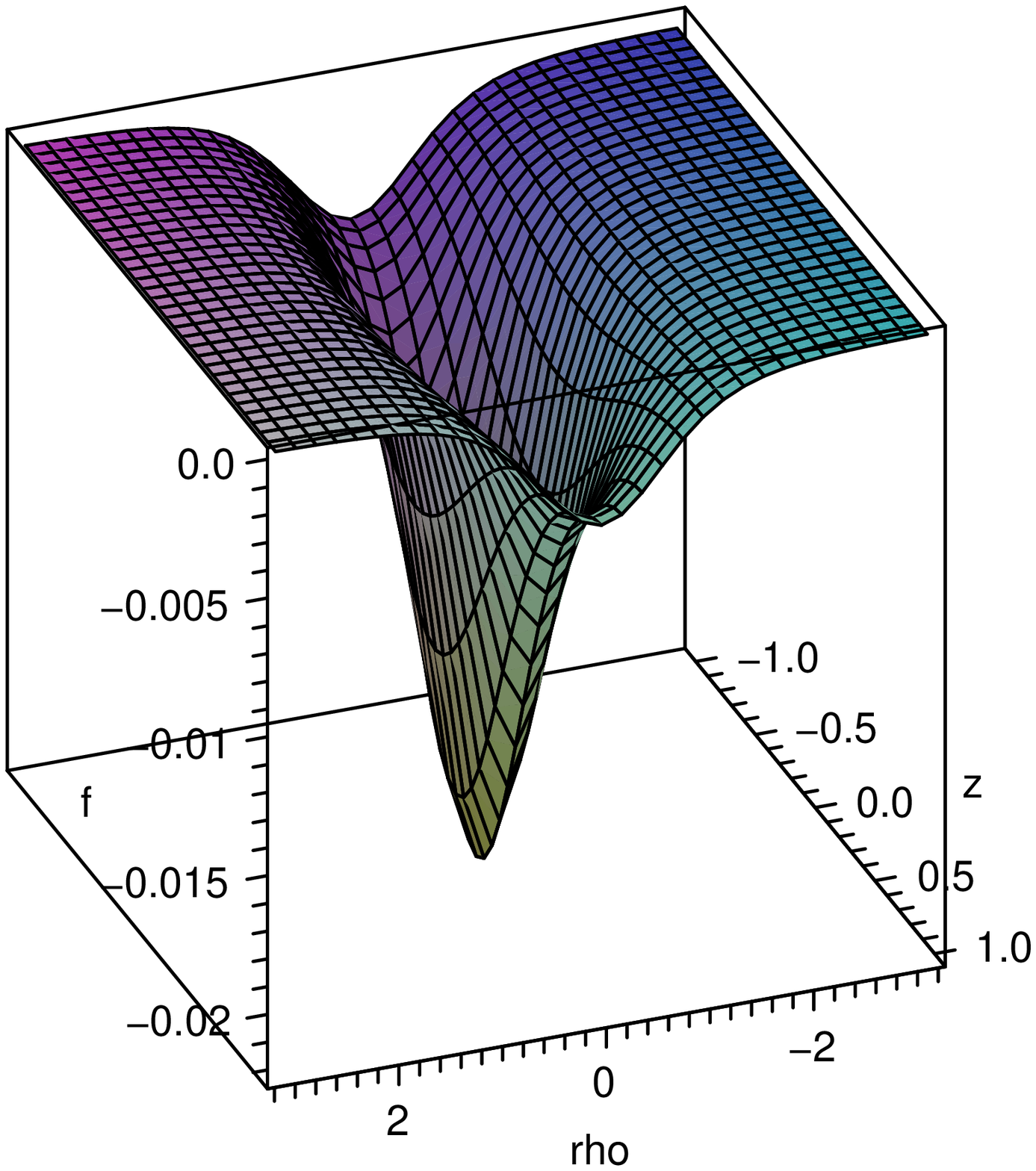} &
 \includegraphics[width=5cm]{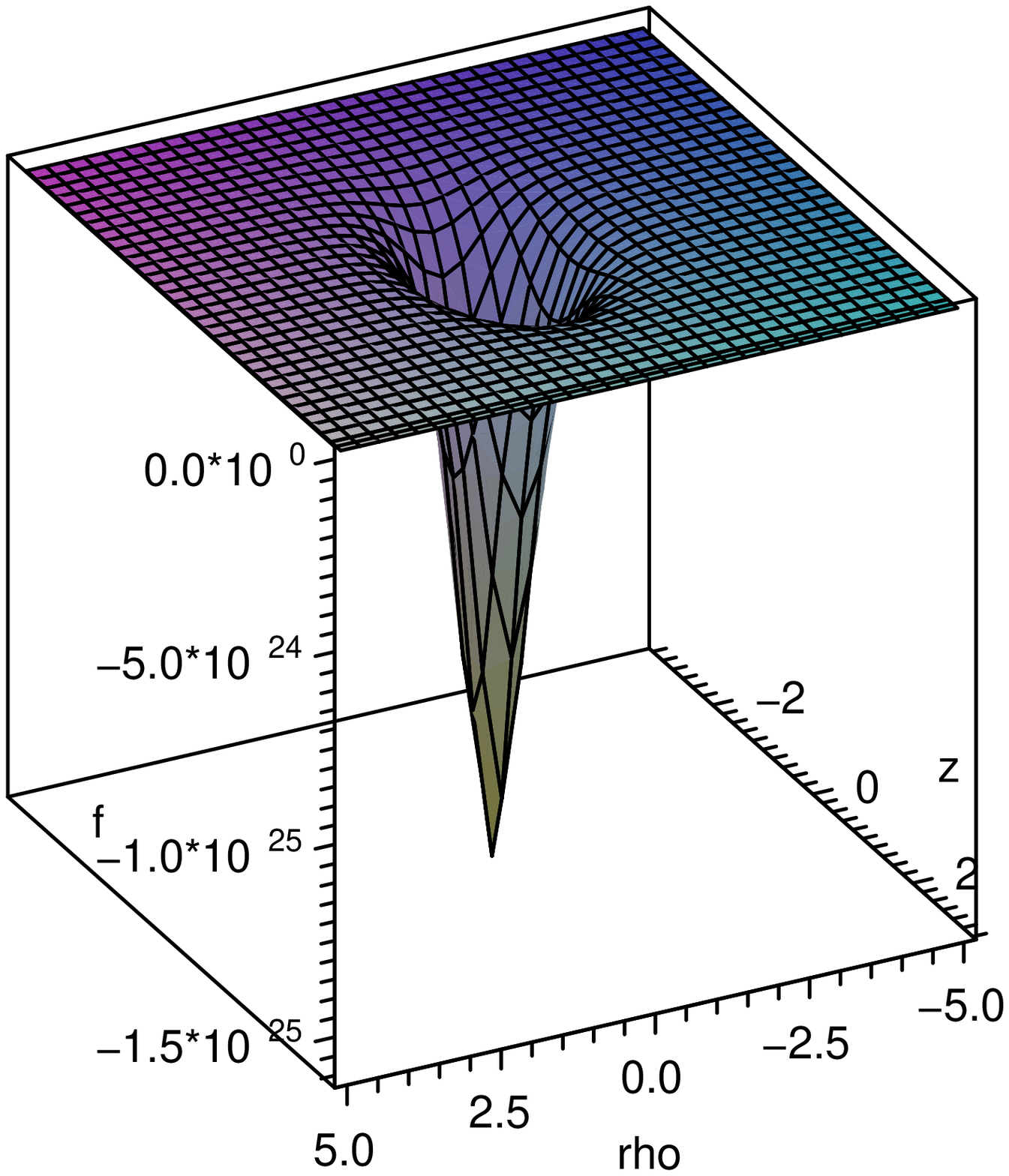} \\
 (a) & (b) & (c) \\
 \end{tabular}
\caption{The harmonic function $-\Rii^6/(2\pi^2 \lp^6 N)f(\rho,z)$
(where $f(\rho,z)$ is given by \reef{M2fsumSc}) for (a) $L/\Rii =
50$, (b) $L/\Rii = 3$, and (c) $L/\Rii = 10^{-4}$. }
\label{fPlots} }

\section{Radial expectation values} \label{except}

In order to compute the expectation value $\langle r\rangle$ (of the
radial coordinate), the radial eigenfunctions must be normalized.
Here, we normalize the modes following the prescription of
ref.~\cite{polchinski}. We briefly review their formalism (which is
a simple generalization of the usual normalization of modes in four
dimensional quantum field theory) before modifying it appropriately
for our uses.

Consider a scalar field propagating in a $(d+1+k)$-dimensional
spacetime with metric
\[
ds^2 (g) = f(y) \eta_{\mu \nu}dx^\mu dx^\nu + g_{\perp mn}dy^m dy^n
\]
where $\mu,\nu =0,...d$, $m,n=1,...,k$.
For a scalar field with the Lagrangian density
\beq
\mathcal{L} = \sqrt{-g} g^{ab} \partial_a \Phi \partial_b \Phi , \labell{lagFree}
\eeq
the canonical commutator is
\[
\left[\Phi({\bf x},y),\dot{\Phi}({\bf x}',y')  \right] =
i \frac{f(y)}{\sqrt{-g}}\delta^d({\bf x}-{\bf x}')\delta^k(y-y').
\]
Specializing for the moment to the case $d=4$ and expanding the
field as
\[
\Phi(x,y) = \sum _\al \int \frac{D^3{\bf k}}{(2\pi )^3}
\frac{1}{2k_0}\left(a_\al({\bf k})e^{ik\cdot x} \psi_\al(y) + h.c. \right)
\]
with the oscillators covariantly normalized as $[a_\al({\bf k}),
a_\beta^\dagger({\bf k}')]=(2\pi)^3\delta^{(3)}({\bf k}-{\bf
k'})\delta_{\al \beta}$, the modes in $y$-space are then normalized
according to
\[
\int d^k y \frac{\sqrt{g}}{f} \psi_\al (y) \psi_\beta^\ast (y) = \delta_{\al \beta}.
\]

For any one of the D$p$-D$k$ brane configurations, the Lagrangian
density for the scalar fields $\chi$ corresponding to fluctuations
of the D$k$-brane transverse to the D$p$-branes was (omitting
multiplicative constants)
\beq
\mathcal{L} = e^{-\phi} \sqrt{-\det g_{ab}} f^{-1} g^ab
\partial_a \chi \partial _b \chi \labell{lagD}
\eeq
where $g_{ab}$ is the induced metric on the D$k$-brane. Comparing
\reef{lagFree} and \reef{lagD} suggests that for the scalar field
confined to the D$k$-brane, the factor $\sqrt{-\det g_{ab}}$ should
be replaced by $e^{-\phi}\sqrt{-\det g_{ab}}f^{-1}$.  Thus, the
canonical commutator should be
\[
\left[\chi({\bf x},y),\dot{\chi}({\bf x}',y')  \right] = i
\frac{f(y)}{e^{-\phi}\sqrt{-\det g_{ab}}f^{-1}}\delta^d({\bf x}-{\bf x}')\delta^k(y-y')
\]
and with $\chi$ expanded in terms of Fourier modes and spherical
harmonics (as described in the text), the measure for the
normalization for the radial functions $\phi(\vr)$ should be
\beq
\mu (\vr) = e^{-\phi}\sqrt{-\det g_{ab}}f^{-2}.
\eeq
Hence, the radial expectation value of any function will be given by
\beq \langle f(\vr) \rangle = \frac{\int _0 ^\infty f(\vr)\, \mu
(\vr) |\phi(\vr)|^2 d\vr}{\int_0^\infty \mu (\vr) |\phi(\vr)|^2
d\vr}. \labell{radExp} \eeq
where the radial measure for any one of the D$p$-D$k$ brane systems is
\[
\mu_{p,k}(\vr)=\frac{\vr^{1+\frac{k-p}{2}}}{(1+\vr^2)^{\frac{7-p}{2}}}.
\]

Given the standard duality between energy and radius, \ie
$U=r/\al'$, the radial profiles give the wave-function in energy ---
that is, $\mu (\vr) |\phi(\vr)|^2$ serves somewhat like a parton
distribution function. The expectation value $\langle r\rangle$
gives a measure of the mean energy of the components of a given
meson state. Using \reef{radExp} and $\phi(\vr)$ for the transverse
scalar fluctuations of the probe brane, we computed $2\pi
\langle U\rangle/m_q= \langle (1+\vr^2)^{1/2} \rangle$ for the D$p$-D$k$ systems. Note that for $p=3$, the calculations can
be done analytically while for $p\ne 3$, the results are obtained
numerically. Explicit results for the D2-D4 and D4-D6 configurations
are displayed in Table \ref{rd2d4}. These are
typical of the general case. Note that $\langle U \rangle$ increases
when the radial quantum number is increased, but decreases with increasing $\ell_2$.

In \cite{spectra}, it was noted for the D3-D7 system that for large
R-charge, $\ell\gg n$, the radial profiles were sharply peaked at
$\vr_{\rm peak}=1$. One can see that the same peaking arises for the
mesons on other probe branes in the D3 background from the analytic
solutions, \eg, eqs.~\reef{abcx1}, \reef{abcx2} and \reef{abcx3}.
This radius corresponds to the radius of a null geodesic on the
induced probe brane metric orbiting on the internal space at a
constant value of $\vr$. One expects a similar peaking for occurs
for general $p$, for which the geodesic analysis yields $\vr_{\rm
peak}^2=2/(5-p)$. For the large R-charge limit, $\ell\gg n$, one can
then infer $2\pi \langle U\rangle/m_q\simeq\sqrt{(7-p)/(5-p)}$.

\end{document}